\documentclass[11pt]{article}
\usepackage[pdftex]{graphicx}
\usepackage{times}
\usepackage{amsmath, amssymb, amstext}
\usepackage{bm}
\usepackage{cancel}
\usepackage[margin=2.5cm]{geometry}
\usepackage[margin=20pt,small,bf]{caption}
\raggedright
\newcommand{\Norm}[1]{\ensuremath{\lVert#1\rVert}}

\newcommand{\Dela}{\ensuremath{\Delta a}}
\newcommand{\Delb}{\ensuremath{\Delta b}}
\newcommand{\Delt}{\ensuremath{\Delta t}}

\newcommand{\Truesdell}[1]{\ensuremath{\overset{\circ}{#1}}}

\newcommand{\Ep}{\ensuremath{\epsilon_p}}

\newcommand{\Epdot}[1]{\ensuremath{\dot{\epsilon}_{#1}}}

\newcommand{\SfA}{\ensuremath{\boldsymbol{\mathsf{A}}}}
\newcommand{\SfC}{\boldsymbol{\mathsf{C}}}

\newcommand{\SfI}{\ensuremath{\boldsymbol{\mathsf{I}}}}

\newcommand{\Balpha}{\ensuremath{\boldsymbol{\alpha}}}
\newcommand{\Bbeta}{\ensuremath{\boldsymbol{\beta}}}

\newcommand{\Bpsi}{\ensuremath{\boldsymbol{\psi}}}

\newcommand{\Bkappa}{\ensuremath{\boldsymbol{\kappa}}}
\newcommand{\Bnabla}{\ensuremath{\boldsymbol{\nabla}}}

\newcommand{\Bsig}{\ensuremath{\boldsymbol{\sigma}}}

\newcommand{\Bvarphi}{\ensuremath{\boldsymbol{\varphi}}}

\newcommand{\Beta}{\ensuremath{\boldsymbol{\eta}}}
\newcommand{\Bone}{\ensuremath{\boldsymbol{\mathit{1}}}}
\newcommand{\Bzero}{\ensuremath{\boldsymbol{\mathit{0}}}}
\newcommand{\BoneHat}{\ensuremath{\widehat{\Bone}}}

\newcommand{\Ba}{\ensuremath{\mathbf{a}}}
\newcommand{\Bb}{\ensuremath{\mathbf{b}}}

\newcommand{\Bd}{\ensuremath{\boldsymbol{d}}}
\newcommand{\Be}{\ensuremath{\mathbf{e}}}

\newcommand{\Bf}{\ensuremath{\mathbf{f}}}
\newcommand{\BfT}{\ensuremath{\boldsymbol{f}}}

\newcommand{\Bl}{\ensuremath{\boldsymbol{l}}}

\newcommand{\Bn}{\ensuremath{\mathbf{n}}}

\newcommand{\Bq}{\ensuremath{\mathbf{q}}}

\newcommand{\Bs}{\ensuremath{\boldsymbol{s}}}
\newcommand{\Bt}{\ensuremath{\mathbf{t}}}
\newcommand{\Bu}{\ensuremath{\mathbf{u}}}
\newcommand{\Bv}{\ensuremath{\mathbf{v}}}
\newcommand{\Bw}{\ensuremath{\mathbf{w}}}
\newcommand{\BwT}{\ensuremath{\boldsymbol{w}}}
\newcommand{\Bx}{\ensuremath{\mathbf{x}}}
\newcommand{\BA}{\ensuremath{\boldsymbol{A}}}
\newcommand{\BB}{\ensuremath{\boldsymbol{B}}}
\newcommand{\BC}{\ensuremath{\boldsymbol{C}}}

\newcommand{\BE}{\ensuremath{\boldsymbol{E}}}
\newcommand{\BEv}{\ensuremath{\mathbf{E}}}
\newcommand{\BF}{\ensuremath{\boldsymbol{F}}}

\newcommand{\BN}{\ensuremath{\boldsymbol{N}}}

\newcommand{\BP}{\ensuremath{\boldsymbol{P}}}

\newcommand{\BQ}{\ensuremath{\boldsymbol{Q}}}
\newcommand{\BR}{\ensuremath{\boldsymbol{R}}}
\newcommand{\BS}{\ensuremath{\boldsymbol{S}}}

\newcommand{\BU}{\ensuremath{\boldsymbol{U}}}

\newcommand{\BX}{\ensuremath{\mathbf{X}}}

\newcommand{\Tor}{\ensuremath{\text{or}}}
\newcommand{\Tr}[1]{\ensuremath{\text{tr}\left(#1\right)}}

\newcommand{\Tint}{\ensuremath{\text{int}}}
\newcommand{\Text}{\ensuremath{\text{ext}}}

\newcommand{\Tbod}{\ensuremath{\text{bod}}}

\newcommand{\Half}{\ensuremath{\frac{1}{2}}}
\newcommand{\Sthr}{\ensuremath{\sqrt{3}}}

\newcommand{\TT}{\ensuremath{\frac{3}{2}}}
\newcommand{\Third}{\ensuremath{\frac{1}{3}}}
\newcommand{\Sixth}{\ensuremath{\frac{1}{6}}}
\newcommand{\Ninth}{\ensuremath{\frac{1}{9}}}

\newcommand{\Bcross}[2]{\ensuremath{#1\boldsymbol{\times}#2}}
\newcommand{\Bdot}[2]{\ensuremath{#1\cdot#2}}
\newcommand{\Dyad}[2]{\ensuremath{#1\boldsymbol{\otimes}#2}}
\newcommand{\Grad}[1]{\ensuremath{\Bnabla #1}}

\newcommand{\Div}[1]{\ensuremath{\Bdot{\Bnabla}{#1}}}

\newcommand{\Gradu}{\ensuremath{\Grad{\Bu}}}

\newcommand{\Gradv}{\ensuremath{\Grad{\Bv}}}
\newcommand{\GradvT}{\ensuremath{(\Gradv)^T}}

\newcommand{\LimDelt}{\ensuremath{\lim_{\Delt\rightarrow 0}}}

\newcommand{\Deriv}[2]{\ensuremath{\frac{d #1}{d #2}}}

\newcommand{\Partial}[2]{\ensuremath{\frac{\partial #1}{\partial #2}}}
\newcommand{\PPartial}[2]{\ensuremath{\frac{\partial^2 #1}{\partial #2^2}}}
\newcommand{\PPartialA}[3]{\ensuremath{\frac{\partial^2 #1}{\partial #2\partial#3}}}

\newcommand{\IntOmega}{\ensuremath{\int_{\Omega}}}
\newcommand{\IntDomega}{\ensuremath{\int_{\Domega}}}
\newcommand{\IntOmegac}{\ensuremath{\int_{\Omega_c}}}
\newcommand{\IntOmegapc}{\ensuremath{\int_{\Omega_p\cap\Omega_c}}}
\newcommand{\IntOmegap}{\ensuremath{\int_{\Omega_p\cap\Omega}}}
\newcommand{\IntOmegaq}{\ensuremath{\int_{\Omega_q\cap\Omega}}}

\newcommand{\IntGammat}{\ensuremath{\int_{\Gamma_t}}}

\newcommand{\IntGammaq}{\ensuremath{\int_{\Gamma_q}}}

\newcommand{\IntGamma}{\ensuremath{\int_{\Gamma}}}
\newcommand{\IntOmegat}{\ensuremath{\int_{\Omega(t)}}}
\newcommand{\IntDomegat}{\ensuremath{\int_{\Domega(t)}}}
\newcommand{\IntOmegatDelt}{\ensuremath{\int_{\Omega(t+\Delt)}}}
\newcommand{\IntOmegar}{\ensuremath{\int_{\Omega_0}}}

\newcommand{\Bart}{\ensuremath{\bar{\mathbf{t}}}}
\newcommand{\BWs}{\ensuremath{\boldsymbol{W}^{*}}}
\newcommand{\BWss}{\ensuremath{\boldsymbol{W}^{**}}}

\newcommand{\BWssq}{\ensuremath{\boldsymbol{W}_q^{**}}}
\newcommand{\BWsDev}{\ensuremath{\text{dev}(\BWs)}}
\newcommand{\BWsVol}{\ensuremath{\Third\text{tr}(\BWs)~\Bone}}
\newcommand{\What}{\ensuremath{\widehat{w}}}
\newcommand{\ThetaDot}{\ensuremath{\dot{\theta}}}
\newcommand{\Sump}{\ensuremath{\sum_{p=1}^{n_p}}}
\newcommand{\Sumpq}{\ensuremath{\sum_{q=1}^{n_p}}}
\newcommand{\Sumpc}{\ensuremath{\sum_{p=1}^{n_p^c}}}
\newcommand{\Sumpcq}{\ensuremath{\sum_{q=1}^{n_p^c}}}
\newcommand{\Sumg}{\ensuremath{\sum_{g=1}^{n_g}}}
\newcommand{\Sumgc}{\ensuremath{\sum_{g=1}^{n_g^c}}}
\newcommand{\Sumgh}{\ensuremath{\sum_{h=1}^{n_g}}}
\newcommand{\Bvdot}{\ensuremath{\dot{\mathbf{v}}}}
\newcommand{\Tdot}{\ensuremath{\dot{T}}}
\newcommand{\LBr}{\ensuremath{\left(}}
\newcommand{\LBc}{\ensuremath{\left\{}}
\newcommand{\LBs}{\ensuremath{\left[}}
\newcommand{\LBn}{\ensuremath{\left.}}
\newcommand{\RBr}{\ensuremath{\right)}}
\newcommand{\RBc}{\ensuremath{\right\}}}
\newcommand{\RBs}{\ensuremath{\right]}}
\newcommand{\RBn}{\ensuremath{\right.}}
\newcommand{\BsHat}{\ensuremath{\widehat{\boldsymbol{s}}}}
\newcommand{\BnHat}{\ensuremath{\widehat{\boldsymbol{n}}}}
\newcommand{\Ss}{\ensuremath{\sigma_s}}
\newcommand{\Sm}{\ensuremath{\sigma_m}}
\newcommand{\Sy}{\ensuremath{\sigma_y}}

\newcommand{\Ssdot}{\ensuremath{\dot{\sigma}_s}}
\newcommand{\Smdot}{\ensuremath{\dot{\sigma}_m}}
\newcommand{\Lambdadot}{\ensuremath{\dot{\lambda}}}
\newcommand{\OSthr}{\ensuremath{\cfrac{1}{\Sthr}}}

\newcommand{\Barq}{\ensuremath{\overline{q}}}
\newcommand{\Domega}{\ensuremath{\partial{\Omega}}}

\newcommand{\DA}{\ensuremath{\text{dA}}}

\newcommand{\DV}{\ensuremath{\text{dV}}}
\newcommand{\BCe}{\ensuremath{\mathcal{E}}}
\newcommand{\GradX}[1]{\ensuremath{\Bnabla_0~#1}}
\newcommand{\DivX}[1]{\ensuremath{\Bdot{\Bnabla_0}{#1}}}

\newcommand{\Etadot}{\ensuremath{\dot{\eta}}}

\newcommand{\BEdot}{\ensuremath{\dot{\BE}}}
\newcommand{\BFdot}{\ensuremath{\dot{\BF}}}

\newcommand{\BSdot}{\ensuremath{\dot{\BS}}}

\newcommand{\Beq}{\begin{equation}}
\newcommand{\Eeq}{\end{equation}}
\newcommand{\Bal}{\begin{aligned}}
\newcommand{\Eal}{\end{aligned}}

\begin{document}

\title{A Material Point Method Formulation for Plasticity}
\author{B. Banerjee\footnote{email: b.banerjee.nz@gmail.com, 
                             current address: 24 Balfour Road, Parnell, Auckland, NZ}\\
        Department of Mechanical Engineering, University of Utah, \\
        Salt Lake City, UT 84112, USA}
\date{March 08, 2007}
\maketitle

\begin{abstract}
This paper discusses a general formulation of the material point method in the context
of additive decomposition rate-independent plasticity.  The process of generating the weak form
shows that volume integration over deforming particles can be major source of error in
the MPM method.  Several useful identities and other results are derived in the appendix.
\end{abstract}

\section{Governing Equations}
  The equations that govern the motion of an elastic-plastic solid 
  include the balance laws for mass, momentum, and energy.  Kinematic equations
  and constitutive relations are needed to complete the system of equations.
  Physical restrictions on the form of the constitutive relations are 
  imposed by an entropy inequality that expresses the second law of 
  thermodynamics in mathematical form.

  The balance laws express the idea that the rate of change of a quantity
  (mass, momentum, energy) in a volume must arise from three causes:
  \begin{enumerate}
    \item the physical quantity itself flows through the surface that bounds
          the volume,
    \item there is a source of the physical quantity on the surface of the
          volume, or/and,
    \item there is a source of the physical quantity inside the volume.
  \end{enumerate}

  Let $\Omega$ be the body (an open subset of Euclidean space) and let 
  $\Domega$ be its surface (the boundary of $\Omega$).  
  
  Let the motion of material points in the body be described by the map
  \[
    \Bx = \Bvarphi(\BX) = \Bx(\BX)
  \]
  where $\BX$ is the position of a point in the initial configuration 
  and $\Bx$ is the location of the same point in the deformed configuration.
  The deformation gradient ($\BF$) is given by
  \[
    \BF = \Partial{\Bx}{\BX} = \GradX{\Bx} ~.
  \]

  \subsection{Balance Laws}
  Let $f(\Bx,t)$ be a physical quantity that is flowing
  through the body.  Let $g(\Bx,t)$ be sources on the surface of the body
  and let $h(\Bx,t)$ be sources inside the body.  Let $\Bn(\Bx,t)$ be the 
  outward unit normal to the surface $\Domega$.  Let $\Bv(\Bx,t)$ be the 
  velocity of the physical particles that carry the physical quantity that is 
  flowing.  Also, let the speed at which the bounding surface $\Domega$ is 
  moving be $u_n$ (in the direction $\Bn$).

  Then, balance laws can be expressed in the general form (\cite{Wright02})
  \[
    \Deriv{}{t}\left[\int_{\Omega} f(\Bx,t)~\DV\right] = 
      \int_{\Domega} f(\Bx,t)[u_n(\Bx,t) - \Bv(\Bx,t)\cdot\Bn(\Bx,t)]~\DA + 
      \int_{\Domega} g(\Bx,t)~\DA + \int_{\Omega} h(\Bx,t)~\DV ~.
  \]
  Note that the functions $f(\Bx,t)$, $g(\Bx,t)$, and $h(\Bx,t)$ can be scalar
  valued, vector valued, or tensor valued - depending on the physical
  quantity that the balance equation deals with.

  It can be shown that the balance laws of mass, momentum, and energy
  can be written as (see Appendix):
  \[
    \begin{aligned}
      \dot{\rho} + \rho~\Div{\Bv} & = 0 
          & & \qquad\text{Balance of Mass} \\
      \rho~\dot{\Bv} - \Div{\Bsig} - \rho~\Bb & = 0 
          & & \qquad\text{Balance of Linear Momentum} \\
      \Bsig & = \Bsig^T
          & & \qquad\text{Balance of Angular Momentum} \\
      \rho~\dot{e} - \Bsig:(\Gradv) + \Div{\Bq} - \rho~s & = 0
          & & \qquad\text{Balance of Energy.}
    \end{aligned}
  \]
  In the above equations $\rho(\Bx,t)$ is the mass density (current), 
  $\dot{\rho}$ is the material time derivative of $\rho$, $\Bv(\Bx,t)$ is the 
  particle velocity, $\dot{\Bv}$ is the material time derivative of $\Bv$, 
  $\Bsig(\Bx,t)$ is the Cauchy stress tensor, $\Bb(\Bx,t)$ is the body force 
  density, $e(\Bx,t)$ is the internal energy per unit mass, $\dot{e}$ is the 
  material time derivative of $e$, $\Bq(\Bx,t)$ is the heat flux vector, 
  and $s(\Bx,t)$ is an energy source per unit mass.
  
  With respect to the reference configuration, the balance laws can be written
  as \footnote{
  We can alternatively define the nominal stress tensor $\BN$ which is the
  transpose of the first Piola-Kirchhoff stress tensor such that
  \[
    \BN := \BP^T = \det(\BF)~(\BF^{-1}\cdot\Bsig)^T = 
        \det(\BF)~\Bsig\cdot\BF^{-T} ~.
  \]
  Then the balance laws become
  \[
    \begin{aligned}
      \rho~\det(\BF) - \rho_0 &= 0 & &  \qquad \text{Balance of Mass} \\
      \rho_0~\ddot{\Bx} - \DivX{\BN} -\rho_0~\Bb & = 0  & & 
        \qquad \text{Balance of Linear Momentum} \\
      \BF\cdot\BN^T & = \BN\cdot\BF^T  & & 
        \qquad \text{Balance of Angular Momentum} \\ 
      \rho_0~\dot{e} - \BN:\dot{\BF} + \DivX{\Bq} - \rho_0~s & = 0
          & & \qquad\text{Balance of Energy.} 
    \end{aligned}
  \]
  }
  \[
    \begin{aligned}
      \rho~\det(\BF) - \rho_0 &= 0 & &  \qquad \text{Balance of Mass} \\
      \rho_0~\ddot{\Bx} - \DivX{\BP^T} -\rho_0~\Bb & = 0  & & 
        \qquad \text{Balance of Linear Momentum} \\
      \BF\cdot\BP & = \BP^T\cdot\BF^T  & & 
        \qquad \text{Balance of Angular Momentum} \\ 
      \rho_0~\dot{e} - \BP^T:\dot{\BF} + \DivX{\Bq} - \rho_0~s & = 0
          & & \qquad\text{Balance of Energy.} 
    \end{aligned}
  \]
  In the above, $\BP$ is the first Piola-Kirchhoff stress tensor, and 
  $\rho_0$ is the mass density in the reference configuration.  The first 
  Piola-Kirchhoff stress tensor is related to the Cauchy stress tensor by
  \[
    \BP = \det(\BF)~\BF^{-1}\cdot\Bsig ~.
  \]
  The gradient and divergence operators are defined such that
  \[
    \Grad{\Bv} = \sum_{i,j = 1}^3 \Partial{v_i}{x_j}\Dyad{\Be_i}{\Be_j} = 
        v_{i,j}\Dyad{\Be_i}{\Be_j} ~;~~
    \Div{\Bv} =  \sum_{i=1}^3 \Partial{v_i}{x_i} = v_{i,i} ~;~~
    \Div{\BS} = \sum_{i,j=1}^3 \Partial{S_{ij}}{x_j}~\Be_i 
          = \sigma_{ij,j}~\Be_i ~.
  \]
  where $\Bv$ is a vector field, $BS$ is a second-order tensor field, and
  $\Be_i$ are the components of an orthonormal basis in the current 
  configuration.  Also,
  \[
    \GradX{\Bv} = \sum_{i,j = 1}^3 \Partial{v_i}{X_j}\Dyad{\BEv_i}{\BEv_j} = 
        v_{i,j}\Dyad{\BEv_i}{\BEv_j} ~;~~
    \DivX{\Bv} =  \sum_{i=1}^3 \Partial{v_i}{X_i} = v_{i,i} ~;~~
    \DivX{\BS} = \sum_{i,j=1}^3 \Partial{S_{ij}}{X_j}~\BEv_i = S_{ij,j}~\BEv_i 
  \]
  where $\Bv$ is a vector field, $BS$ is a second-order tensor field, and
  $\BEv_i$ are the components of an orthonormal basis in the reference 
  configuration.

  The contraction operation is defined as
  \[
    \BA:\BB = \sum_{i,j=1}^3 A_{ij}~B_{ij} = A_{ij}~B_{ij} ~.
  \]

  \subsection{The Clausius-Duhem Inequality}
  The Clausius-Duhem inequality can be used to express the second law of
  thermodynamics for elastic-plastic materials.  This inequality is a 
  statement concerning the irreversibility of natural processes, especially
  when energy dissipation is involved.

  Just like in the balance laws in the previous section, we assume that
  there is a flux of a quantity, a source of the quantity, and an internal
  density of the quantity per unit mass.  The quantity of interest in this
  case is the entropy.  Thus, we assume that there is an entropy flux, an
  entropy source, and an internal entropy density per unit mass ($\eta$)
  in the region of interest.

  Let $\Omega$ be such a region and let $\Domega$ be its boundary.  Then
  the second law of thermodynamics states that the rate of increase of
  $\eta$ in this region is greater than or equal to the sum of that supplied 
  to $\Omega$ (as a flux or from internal sources) and the change of the 
  internal entropy density due to material flowing in and out of the region.
  
  Let $\Domega$ move with a velocity $u_n$ and let particles inside 
  $\Omega$ have velocities $\Bv$.  Let $\Bn$ be the unit outward normal 
  to the surface $\Domega$.  Let $\rho$ be the density of matter in the 
  region, $\bar{q}$ be the entropy flux at the surface, and $r$ be the 
  entropy source per unit mass.  
  Then (see \cite{Wright02,Batra06}), the entropy inequality may be written as
  \[
    \Deriv{}{t}\left(\IntOmega \rho~\eta~\DV\right) \ge
    \IntDomega \rho~\eta~(u_n - \Bv\cdot\Bn)~\DA + 
    \IntDomega \bar{q}~\DA + \IntOmega \rho~r~\DV ~.
  \]
  The scalar entropy flux can be related to the vector flux at the surface
  by the relation $\bar{q} = -\Bpsi(\Bx)\cdot\Bn$.  Under the assumption 
  of incrementally isothermal conditions (see \cite{Maugin99} for a detailed
  discussion of the assumptions involved), we have
  \[
    \Bpsi(\Bx) = \cfrac{\Bq(\Bx)}{T} ~;~~ r = \cfrac{s}{T}
  \]
  where $\Bq$ is the heat flux vector, $s$ is a energy source per unit mass, 
  and $T$ is the absolute temperature of a material point at $\Bx$ at time $t$.

  We then have the Clausius-Duhem inequality in integral form:
  \[
    \Deriv{}{t}\left(\IntOmega \rho~\eta~\DV\right) \ge
    \IntDomega \rho~\eta~(u_n - \Bv\cdot\Bn)~\DA - 
    \IntDomega \cfrac{\Bq\cdot\Bn}{T}~\DA + \IntOmega \cfrac{\rho~s}{T}~\DV ~.
  \]
  We can show that (see Appendix) the entropy inequality may be 
  written in differential form as
  \[
    \rho~\dot{\eta} \ge - \Div{\left(\cfrac{\Bq}{T}\right)} 
       + \cfrac{\rho~s}{T} ~.
  \]
  In terms of the Cauchy stress and the internal energy, the Clausius-Duhem
  inequality may be written as (see Appendix)
  \[
      \rho~(\dot{e} - T~\dot{\eta}) - \Bsig:\Gradv \le 
           - \cfrac{\Bq\cdot\Grad{T}}{T} ~.
  \]
  
  \subsection{Constitutive Relations}
  A set of constitutive equations is required to close to system of balance
  laws.  For large deformation plasticity, we have to define appropriate 
  kinematic quantities and stress measures so that constitutive relations
  between them may have a physical meaning.

  \subsubsection{Thermoelastic Relations}
  Let the fundamental kinematic quantity be the deformation gradient ($\BF$)
  which is given by
  \[
    \BF = \Partial{\Bx}{\BX} = \GradX{\Bx} ~;~~ \det\BF > 0 ~.
  \]
  A thermoelastic material is one in which the internal energy ($e$) is a 
  function only of $\BF$ and the specific entropy ($\eta$), that is\footnote{
  If a thermoelastic body is subjected to a rigid body rotation $\BQ$, then
  its internal energy should not change.  After a rotation, the new
  deformation gradient ($\hat{\BF}$) is given by
  \[
     \hat{\BF} = \BQ\cdot\BF ~.
  \]
  Since the internal energy does not change, we must have
  \[
     e = \bar{e}(\hat{\BF}, \eta) = \bar{e}(\BF, \eta) ~.
  \]
  Now, from the polar decomposition theorem,  $\BF = \BR\cdot\BU$ where $\BR$ is
  the orthogonal rotation tensor (i.e., $\BR\cdot\BR^T = \BR^T\cdot\BR = \Bone$)
  and $\BU$ is the symmetric right stretch tensor.  Therefore, 
  \[
     \bar{e}(\BQ\cdot\BR\cdot\BU, \eta) = \bar{e}(\BF, \eta) ~.
  \]
  Now, we can choose any rotation $\BQ$.  In particular, if we choose 
  $\BQ = \BR^T$, we have
  \[
     \bar{e}(\BR^T\cdot\BR\cdot\BU, \eta) = 
       \bar{e}(\Bone\cdot\BU, \eta) = \tilde{e}(\BU, \eta)~.
  \]
  Therefore,
  \[
     \bar{e}(\BU, \eta) = \bar{e}(\BF, \eta) ~.
  \]
  This means that {\em the internal energy depends only on the stretch $\BU$ and
  not on the orientation of the body}.
  }
  \[
     e = \bar{e}(\BF, \eta) ~.
  \]
  For a thermoelastic material, we can show that the 
  entropy inequality can be written as (see Appendix)
  \[
      \rho~\left(\Partial{\bar{e}}{\eta} - T\right)~\dot{\eta} +
              \left(\rho~\Partial{\bar{e}}{\BF} - 
              \Bsig\cdot\BF^{-T}\right):\BFdot 
       + \cfrac{\Bq\cdot\Grad{T}}{T} \le 0 ~.
  \]
  At this stage, we make the following constitutive assumptions:
  \begin{enumerate}
    \item
    Like the internal energy, we assume that $\Bsig$ and $T$ are also functions
    only of $\BF$ and $\eta$, i.e., 
    \[
       \Bsig = \Bsig(\BF, \eta) ~;~~ T = T(\BF, \eta) ~.
    \]
    \item
    The heat flux $\Bq$ satisfies the thermal conductivity inequality and 
    if $\Bq$ is independent of $\dot{\eta}$ and $\BFdot$, we have
    \[
       \Bq\cdot\Grad{T} \le 0  \qquad\implies\qquad 
       -(\Bkappa\cdot\Grad{T})\cdot\Grad{T} \le 0 \qquad\implies\qquad 
       \Bkappa\ge\Bzero
    \]
    i.e., the thermal conductivity $\Bkappa$ is positive semidefinite.
  \end{enumerate}
  Therefore, the entropy inequality may be written as
  \[
    \rho~\left(\Partial{\bar{e}}{\eta} - T\right)~\dot{\eta} +
         \left(\rho~\Partial{\bar{e}}{\BF} - \Bsig\cdot\BF^{-T}\right):\BFdot 
    \le 0 ~.
  \]
  Since $\dot{\eta}$ and $\BFdot$ are arbitrary, the entropy inequality will
  be satisfied if and only if
  \[
    \Partial{\bar{e}}{\eta} - T = 0 \implies
     T = \Partial{\bar{e}}{\eta}
    \qquad\text{and}\qquad
    \rho~\Partial{\bar{e}}{\BF} - \Bsig\cdot\BF^{-T} = \Bzero \implies
    \Bsig = \rho~\Partial{\bar{e}}{\BF}\cdot\BF^T ~.
  \]
  Therefore, 
  \[
    T = \Partial{\bar{e}}{\eta}
    \qquad\text{and}\qquad
    \Bsig = \rho~\Partial{\bar{e}}{\BF}\cdot\BF^T ~.
  \]
  Given the above relations, the energy equation may expressed in terms of 
  the specific entropy as (see Appendix)
  \[
      \rho~T~\dot{\eta} = - \Div{\Bq} + \rho~s ~.
  \]

  \subsubsection{Alternative strain and stress measures}
  The internal energy depends on $\BF$ only through the stretch $\BU$.  A
  strain measure that reflects this fact and also vanishes in the reference
  configuration is the Green strain
  \[
    \BE = \Half(\BF^T\cdot\BF - \Bone) = \Half(\BU^2 - \Bone) ~.
  \]
  Recall that the Cauchy stress is given by
  \[
    \Bsig = \rho~\Partial{\bar{e}}{\BF}\cdot\BF^T ~.
  \]
  We can show that the Cauchy stress can be expressed in terms of the 
  Green strain as (see Appendix)
  \[
    \Bsig = \rho~\BF\cdot\Partial{\bar{e}}{\BE}\cdot\BF^T ~.
  \]
  Recall that the nominal stress tensor is defined as
  \[
     \BN = \det\BF~(\Bsig\cdot\BF^{-T})~.
  \]
  From the conservation of mass, we have $\rho_0 = \rho~\det\BF$.  Hence,
  \[
     \BN = \cfrac{\rho_0}{\rho}~\Bsig\cdot\BF^{-T}~.
  \]
  The nominal stress is unsymmetric.  We can define a symmetric stress 
  measure with respect to the reference configuration call the
  second Piola-Kirchhoff stress tensor ($\BS$):
  \[
     \BS := \BF^{-1}\cdot\BN = \BP\cdot\BF^{-T} = 
       \cfrac{\rho_0}{\rho}~\BF^{-1}\cdot\Bsig\cdot\BF^{-T}~.
  \]
  In terms of the derivatives of the internal energy, we have
  \[
     \BS = \cfrac{\rho_0}{\rho}~\BF^{-1}\cdot
      \left(\rho~\BF\cdot\Partial{\bar{e}}{\BE}\cdot\BF^T\right)\cdot\BF^{-T}
      = \rho_0~\Partial{\bar{e}}{\BE} 
  \]
  and
  \[
     \BN = \cfrac{\rho_0}{\rho}~
       \left(\rho~\BF\cdot\Partial{\bar{e}}{\BE}\cdot\BF^T\right)\cdot\BF^{-T} 
       = \rho_0~\BF\cdot\Partial{\bar{e}}{\BE} ~.
  \]
  That is,
  \[
     \BS = \rho_0~\Partial{\bar{e}}{\BE} 
     \qquad\text{and}\qquad
     \BN = \rho_0~\BF\cdot\Partial{\bar{e}}{\BE} \,.
  \]

  \subsubsection{Stress Power}
  The stress power per unit volume is given by $\Bsig:\Gradv$.  In terms of
  the stress measures in the reference configuration, we have
  \[
    \Bsig:\Gradv = 
    \left(\rho~\BF\cdot\Partial{\bar{e}}{\BE}\cdot\BF^T\right):
    (\BFdot\cdot\BF^{-1})
    ~.
  \]
  Using the identity $\BA:(\BB\cdot\BC) = (\BA\cdot\BC^T):\BB$, we have
  \[
    \Bsig:\Gradv = 
    \left[
    \left(\rho~\BF\cdot\Partial{\bar{e}}{\BE}\cdot\BF^T\right)\cdot\BF^{-T}
     \right] :\BFdot
    = \rho~\left(\BF\cdot\Partial{\bar{e}}{\BE}\right):\BFdot 
    = \cfrac{\rho}{\rho_0}~\BN:\BFdot ~.
  \]
  We can alternatively express the stress power in terms of $\BS$ and
  $\dot{\BE}$.  Taking the material time derivative of $\BE$ we have
  \[
     \dot{\BE} = \Half(\dot{\BF^T}\cdot\BF + \BF^T\cdot\BFdot) ~.
  \]
  Therefore, 
  \[ 
     \BS:\dot{\BE} = \Half[\BS:(\dot{\BF^T}\cdot\BF) + \BS:(\BF^T\cdot\BFdot)]~.
  \]
  Using the identities $\BA:(\BB\cdot\BC) = (\BA\cdot\BC^T):\BB
      = (\BB^T\cdot\BA):\BC$ and $\BA:\BB = \BA^T:\BB^T$ and using the 
  symmetry of $\BS$, we have
  \[ 
     \BS:\dot{\BE} = 
       \Half[(\BS\cdot\BF^T):\BFdot^T + (\BF\cdot\BS):\BFdot] = 
       \Half[(\BF\cdot\BS^T):\BFdot + (\BF\cdot\BS):\BFdot] = 
       (\BF\cdot\BS):\BFdot ~.
  \]
  Now, $\BS = \BF^{-1}\cdot\BN$.  Therefore, $\BS:\dot{\BE} = \BN:\BFdot$.
  Hence, the stress power can be expressed as
  \[
      \Bsig:\Gradv = \BN:\BFdot = \BS:\dot{\BE} ~.
  \]
  If we split the velocity gradient into symmetric and skew parts using
  \[
     \Gradv = \Bl = \Bd + \BwT
  \]
  where $\Bd$ is the rate of deformation tensor and $\BwT$ is the spin tensor,
  we have
  \[
     \Bsig:\Gradv = \Bsig:\Bd + \Bsig:\BwT = 
       \Tr{\Bsig^T\cdot\Bd} + \Tr{\Bsig^T\cdot\BwT} =
       \Tr{\Bsig\cdot\Bd} + \Tr{\Bsig\cdot\BwT} ~.
  \]
  Since $\Bsig$ is symmetric and $\BwT$ is skew, we have
  $\Tr{\Bsig\cdot\BwT} = 0$.
  Therefore, $\Bsig:\Gradv = \Tr{\Bsig\cdot\Bd}$.  Hence, we may also
  express the stress power as
  \[
      \Tr{\Bsig\cdot\Bd} = \Tr{\BN^T\cdot\BFdot} = \Tr{\BS\cdot\dot{\BE}} ~.
  \]
    
  \subsubsection{Helmholtz and Gibbs free energy}
  Recall that
  \[
     \BS = \rho_0~\Partial{\bar{e}}{\BE} ~.
  \]
  Therefore,
  \[
     \Partial{\bar{e}}{\BE} = \cfrac{1}{\rho_0}~\BS ~.
  \]
  Also recall that
  \[
     \Partial{\bar{e}}{\eta} = T ~.
  \]
  Now, the internal energy $e = \bar{e}(\BE, \eta)$ is a function only of the 
  Green strain and the specific entropy.  Let us assume, that the above
  relations can be uniquely inverted locally at a material point so that we have
  \[
     \BE = \tilde{\BE}(\BS, T) \qquad\text{and}\qquad 
     \eta = \tilde{\eta}(\BS, T) ~.
  \]
  Then the specific internal energy, the specific entropy, and the stress 
  can also be expressed as functions of $\BS$ and $T$, or $\BE$ and $T$, i.e.,
  \[
    e = \bar{e}(\BE, \eta) = \tilde{e}(\BS, T) = \hat{e}(\BE, T)~;
    \qquad
    \eta = \tilde{\eta}(\BS,T) = \hat{\eta}(\BE,T) ~;
    \qquad\text{and}\qquad \BS = \hat{\BS}(\BE,T)
  \]
  We can show that (see Appendix)
  \[
    \Deriv{}{t}(e - T~\eta) = - \dot{T}~\eta + \cfrac{1}{\rho_0}~\BS:\BEdot
    \qquad\Tor\qquad
    \Deriv{\psi}{t} = - \dot{T}~\eta + \cfrac{1}{\rho_0}~\BS:\BEdot ~.
  \]
  and
  \[
    \Deriv{}{t}(e - T~\eta - \cfrac{1}{\rho_0}~\BS:\BE) = 
      - \dot{T}~\eta - \cfrac{1}{\rho_0}~\BSdot:\BE 
    \qquad\Tor\qquad
    \Deriv{g}{t} = \dot{T}~\eta + \cfrac{1}{\rho_0}~\BSdot:\BE ~.
  \]
  We define the {\bf Helmholtz free energy} as
  \[
      \psi = \hat{\psi}(\BE, T) := e - T~\eta ~.
  \]
  We define the {\bf Gibbs free energy} as
  \[
      g  = \tilde{g}(\BS, T) := - e + T~\eta + \cfrac{1}{\rho_0}~\BS:\BE ~.
  \]
  The functions $\hat{\psi}(\BE,T)$ and $\tilde{g}(\BS,T)$ are unique.  Using 
  these definitions it can be showed that (see Appendix)
  \[
    \Partial{\hat{\psi}}{\BE}  = \cfrac{1}{\rho_0}~\hat{\BS}(\BE,T) ~;~~
    \Partial{\hat{\psi}}{T}  = -\hat{\eta}(\BE,T) ~;~~
    \Partial{\tilde{g}}{\BS}  = \cfrac{1}{\rho_0}~\tilde{\BE}(\BS, T) ~;~~
    \Partial{\tilde{g}}{T}  = \tilde{\eta}(\BS, T) 
  \]
  and
  \[
        \Partial{\hat{\BS}}{T} = - \rho_0~\Partial{\hat{\eta}}{\BE} 
        \qquad\text{and}\qquad
        \Partial{\tilde{\BE}}{T} = \rho_0~\Partial{\tilde{\eta}}{\BS} ~.
  \]

  \subsubsection{Specific Heats}
  The {\bf specific heat at constant strain} (or constant volume) is defined as
  \[
     C_v := \Partial{\hat{e}(\BE,T)}{T} ~.
  \]
  The {\bf specific heat at constant stress} (or constant pressure) is 
  defined as
  \[
     C_p := \Partial{\tilde{e}(\BS, T)}{T} ~.
  \]
  We can show that (see Appendix)
  \[
     C_v =  T~\Partial{\hat{\eta}}{T} = -T~\PPartial{\hat{\psi}}{T} 
  \]
  and
  \[
     C_p = 
       T~\Partial{\tilde{\eta}}{T} + 
       \cfrac{1}{\rho_0}~\BS:\Partial{\tilde{\BE}}{T} = 
       T~\PPartial{\tilde{g}}{T}  
       + \BS:\PPartialA{\tilde{g}}{\BS}{T}~.
  \]
  Also the equation for the balance of energy can be expressed in terms
  of the specific heats as (see Appendix)
  \[
    \begin{aligned}
      \rho~C_v~\Tdot & = \Div{(\Bkappa\cdot\Grad{T})} + \rho~s 
           +\cfrac{\rho}{\rho_0}~T~\Bbeta_S:\BEdot \\
      \rho~\left(C_p - \cfrac{1}{\rho_0}~\BS:\Balpha_E\right)
        ~\Tdot 
         & = \Div{(\Bkappa\cdot\Grad{T})} + \rho~s 
           -\cfrac{\rho}{\rho_0}~T~\Balpha_E:\BSdot  
    \end{aligned}
  \]
  where 
  \[
    \Bbeta_S := \Partial{\hat{\BS}}{T} 
    \qquad \text{and} \qquad
    \Balpha_E := \Partial{\tilde{\BE}}{T} ~.
  \]
  The quantity $\Bbeta_S$ is called the {\bf coefficient of thermal stress}
  and the quantity $\Balpha_E$ is called the {\bf coefficient of thermal 
  expansion}.

  The difference between $C_p$ and $C_v$ can be expressed as
  \[
    C_p - C_v = \cfrac{1}{\rho_0}\left(\BS-T~\Partial{\hat{\BS}}{T}\right):
           \Partial{\tilde{\BE}}{T} ~.
  \]
  However, it is more common to express the above relation in terms of the 
  elastic modulus tensor as (see Appendix for proof)
  \[
    C_p - C_v = \cfrac{1}{\rho_0}~\BS:\Balpha_E + 
           \cfrac{T}{\rho_0}~\Balpha_E:\SfC:\Balpha_E
  \]
  where the {\bf fourth-order tensor of elastic moduli} is defined as
  \[
     \SfC := \Partial{\hat{\BS}}{\tilde{\BE}}
           = \rho_0~\PPartialA{\hat{\psi}}{\tilde{\BE}}{\tilde{\BE}} ~.
  \]
  For isotropic materials with a constant coefficient of thermal expansion
  that follow the St. Venant-Kirchhoff material model, we can show that
  (see Appendix)
  \[
     C_p - C_v = \cfrac{1}{\rho_0}\left[\alpha~\Tr{\BS} + 
           9~\alpha^2~K~T\right]~.
  \]

  \subsubsection{Kinematic Relations}
  %\paragraph{Multiplicative split of deformation gradient tensor:}
  \paragraph{Additive split of the rate of deformation tensor:}

  The rate of deformation tensor ($\Bd$) is given by
  \[
    \Bd  = \Half\LBs\Gradv + \GradvT\RBs
  \]
  where $\Bv$ is the velocity.  We assume that the rate of deformation 
  can be additively decomposed into an elastic part, a plastic part, and
  a thermal part:
  \[
    \Bd = \Bd^e + \Bd^p + \Bd^{th}~.
  \]
  The thermal part of the rate of deformation is computed separately and
  subtracted from $\Bd$ to give
  \begin{equation*}
    \widetilde{\Bd} := \Bd - \Bd^{th} = \Be^e + \Bd^p ~.
  \end{equation*}
  For simplicity, we use the symbol $\Bd$ instead of $\widetilde{\Bd}$ in
  the following development.

  We split the rate of deformation tensors into volumetric and deviatoric
  parts:
  \[
    \Bd  = \Third~\ThetaDot~\Bone + \Beta ~;~~~
    \Bd^e  = \Third~\Tr{\Bd^e}~\Bone + \Beta^e ~;~~~
    \Bd^p  = \Third~\Tr{\Bd^p}~\Bone + \Beta^p
  \]
  where $\ThetaDot = \Tr{\Bd}$, the deviatoric part of the rate of deformation
  tensor is $\Beta = \Bd - \Third~\Tr{\Bd}~\Bone$,
  and $\Bone$ is the second-order identity tensor.

  \subsubsection{Stress}
  To deal with nearly incompressible behavior, we introduce an isomorphic
  split of the Cauchy stress tensor into a volumetric and a deviatoric part
  of the form (\cite{Brannon00}):
  \begin{equation} \label{eq:sigDecomp}
    \Bsig = \Sm \BoneHat + \Ss \BsHat 
    \qquad \Longleftrightarrow \qquad
    \Bsig = p~\Bone + \Bs
  \end{equation}
  where
  \[
    \begin{aligned}
    \Sm & = \OSthr~\Tr{\Bsig} = \Sthr~p ~;~~
    \Ss = \Norm{\Bs} ~;~~
    p = \Third~\Tr{\Bsig} ~;~~
    \Bs = \Bsig - \Third~\Tr{\Bsig}~\Bone \\
    \BoneHat & = \cfrac{\Bone}{\Norm{\Bone}} ~;~~
    \BsHat = \cfrac{\Bs}{\Norm{\Bs}} ~;~~
    \Norm{\BA} = \sqrt{\BA:\BA} ~.
    \end{aligned}
  \]
  Note that the following identities hold:
  \[
    \Sm = \Bsig:\BoneHat ~;~~ \Ss = \Bsig:\BsHat ~;~~
    \BoneHat:\BoneHat = 1 ~;~~ \BsHat:\BsHat = 1 ~;~~ \BoneHat:\BsHat = 0 ~;~~
    \BoneHat : \dot{\BsHat} = 0 ~;~~ \BsHat : \dot{\BsHat} = 0 ~.
  \]
  
  The rate form of equation (\ref{eq:sigDecomp}) is
  \begin{equation}\label{eq:threeterm}
    \dot{\Bsig} = \Smdot~\BoneHat + \Ssdot~\BsHat + \Ss~\dot{\BsHat}
    \qquad \Longleftrightarrow \qquad
    \dot{\Bsig} = \dot{p}~\Bone + \dot{\Bs} ~.
  \end{equation}

  In the following development, we enforce frame indifference while evaluating
  the constitutive relation by assuming that the stresses and rate of 
  deformation have been rotated to the reference configuration using
  \begin{equation*}
    \Bs = \BR^T \cdot \Bs \cdot \BR ~;~~
    \Beta = \BR^T \cdot \Beta \cdot \BR 
  \end{equation*}
  where the rotation tensor $\BR$ is obtained from a polar decomposition of
  the deformation gradient.

\subsubsection{Elastic Relations:}
The constitutive model is assumed to consist of two parts.  The first 
is a hypoelastic model for the deviatoric part of the stress and the 
second is a equation of state for the mean stress.  

The deviatoric stress is given by the rate equation
\[
  \dot{\Bs} = \SfC(\Sm, T):\Beta^e
  \qquad \Longleftrightarrow \qquad
  \dot{\Bs} = \widehat{\SfC}(p, T):\Beta^e
\]
where $\dot{\Bs}$ is the deviatoric stress rate, $\SfC$ 
is a pressure and temperature dependent fourth order elastic modulus tensor, 
and $\Beta^e$ is the deviatoric part of $\Bd^e$ (the elastic part of the 
rate of deformation tensor).  

Let us assume that the deviatoric elastic response of the material is 
isotropic, i.e., \footnote{
Note that since the shear modulus depends on $\Sm$, the rate of $\Bs$ should
have a contribution that contains $\Smdot$.  We ignore this contribution
in the subsequent development.
}
\begin{equation}\label{eq:Bsdot}
  \dot{\Bs} = 2~\mu(\Sm, T)~\Beta^e 
  \qquad \Longleftrightarrow \qquad
  \dot{\Bs} = 2~\widehat{\mu}(p, T)~\Beta^e ~.
\end{equation}

The mean stress ($\Sm$) is calculated using a 
Mie-Gr{\"u}neisen equation of state of the form (\cite{Wilkins99,Zocher00})
\begin{equation} \label{eq:EOSMG}
  \Sm = \Sthr~p = - \frac{\Sthr~\rho_0~C_0^2~(1 - J^e)
           [1 - \Gamma_0 (1 - J^e)]}
           {[1 - S_{\alpha}(1 - J^e)]^2} - 2\Sthr~\Gamma_0~e 
  ~;~~~ J^e := \det{\BF^e} 
\end{equation}
where $C_0$ is the bulk speed of sound, $\rho_0$ is the initial mass density,
$2~\Gamma_0$ is the Gr{\"u}neisen's gamma at the reference state,
$S_{\alpha} = dU_s/dU_p$ is a linear Hugoniot slope coefficient,
$U_s$ is the shock wave velocity, $U_p$ is the particle velocity, and
$e$ is the internal energy density (per unit reference volume), $\BF^e$ is
the elastic part of the deformation gradient.  For isochoric plasticity,
\begin{equation*}
  J^e = J = \det(\BF) = \cfrac{\rho_0}{\rho} ~.
\end{equation*}

The change in internal energy is computed using
\[
  e = \rho_0 \int C_v dT \approx \rho_0~[C_v(T)~T- C_v(T_0)~T_0]
\]
where $T_0$ is the reference temperature and $C_v$ is the specific heat at 
constant volume.  We assume that $C_p$ and $C_v$ are equal.

The rate equation (\ref{eq:threeterm}) has three rate terms.  We need
constitutive relations for each of these three terms.

The rate form of the pressure equation is obtained by taking the material
time derivative of (\ref{eq:EOSMG}) to get
\[
  \Smdot = \Sthr~\dot{p} = 
    \cfrac{\Sthr~\rho_0~C_0^2~J^e~[1 +(S_{\alpha} - 2~\Gamma_0)(1-J^e)]}
                   {[1 - S_{\alpha}(1-J^e)]^3}~\LBr\cfrac{\dot{J^e}}{J^e}\RBr
            - 2\Sthr~\Gamma_0~\dot{e}
\]
Now, 
\begin{equation*}
  \cfrac{\dot{J^e}}{J^e} = \Tr{\Bd^e} = \dot{\theta^e} 
    = \Bd^e:\Bone = \Sthr~\Bd^e:\BoneHat ~;~~ 
  \dot{e} = \rho_0~C_v(T)~\dot{T} ~.
\end{equation*}
For isochoric plasticity, the above relations become
\begin{equation*}
  \cfrac{\dot{J}}{J} = \Tr{\Bd^e} = \Tr{\Bd} = \dot{\theta}
    = \Bd:\Bone = \Sthr~\Bd:\BoneHat ~.
\end{equation*}
Hence,
\[
  \Smdot = \Sthr~\dot{p} = 3~K(J^e)~\Bd^e:\BoneHat
            - 2\Sthr~\rho_0~C_v~\Gamma_0~\dot{T}
\]
where
\[
  K(J^e) = \cfrac{\rho_0~C_0^2~J^e~[1 +(S_{\alpha} - 2~\Gamma_0)(1-J^e)]}
                 {[1 - S_{\alpha}(1-J^e)]^3} ~.
\]
The stress rate $\Ssdot$ is given by
\begin{equation*}
  \Ssdot = \Partial{}{t}(\sqrt{\Bs:\Bs}) = \dot{\Bs}:\BsHat
         = 2~\mu(\Sm, T)~\Beta^e:\BsHat ~.
\end{equation*}
Therefore, 
\[
  \Ssdot = 2~\mu(\Sm, T)~\Beta^e:\BsHat ~.
\]
The rate of $\BsHat$ is obtained by noting that
\begin{equation*} 
  \dot{\BsHat} = \cfrac{\dot{\Bs}}{\Norm{\Bs}} - 
    \LBr \cfrac{\dot{\Bs}:\BsHat}{\Norm{\Bs}} \RBr \BsHat ~.
\end{equation*} 
Therefore,
\[
  \dot{\BsHat} = \cfrac{2~\mu(\Sm,T)}{\Ss}~
                  \LBs \Beta^e - (\Beta^e:\BsHat)\BsHat\RBs ~.
\]
A slightly different, but equally valid, decomposition of the stress rate
can be obtained by setting 
\[
  \Ssdot = 2~\mu(\Sm, T)~\Bd^e:\BsHat ~.
\]
and
\[
  \dot{\BsHat} = \cfrac{2~\mu(\Sm,T)}{\Ss}~
                  \LBs \Bd^e - (\Bd^e:\BoneHat)\BoneHat
                  - (\Bd^e:\BsHat)\BsHat\RBs ~.
\]
We use the above decomposition in the following development.

\subsubsection{Plastic Relations:}
We consider yield functions of the form
\[
  f(\Sm,~\Ss,~\Epdot,~\Ep,~T) \le 0
\]
where $\Epdot{}$ is the {\em total} strain rate, $\Ep$ is the plastic strain,
and $T$ is the temperature.  Here the strain rates and the plastic strain
are defined as
\[
  \Epdot{} := \sqrt{\cfrac{2}{3}~ \Bd:\Bd} ~;~~
  \Epdot{p} := \sqrt{\cfrac{2}{3}~ \Bd^p:\Bd^p} ~;~~
  \Ep := \int_0^t \Epdot{p}~dt^{'}  ~.
\]
As a particular case, we consider a class of 
isotropic yield functions of the von Mises form \footnote{
This form of the yield function can be used for deformation induced 
anisotropic plasticity by converting it to the form used by Maudlin 
and Schiferl (\cite{Maudlin96}).  In addition, the effect of porosity can 
be incorporated using Brannon's approach (\cite{Brannon00}).
}
\begin{equation} \label{eq:yieldFn}
  f(\Sm,~\Ss,~\Epdot,~\Ep,~T) = 
    \TT~\Ss^2 - \Sy^2(\Epdot,~\Ep,~T)~\cfrac{\mu^2(\Sm, ~T)}{\mu_0^2}
\end{equation}
where $\Sy$ is the yield stress, $\mu$ is the shear modulus, and $\mu_0$ is
a reference shear modulus.

At yield, we have
\begin{equation*}
  \TT~\Ss^2 - \Sy^2(\Epdot,~\Ep,~T)~\cfrac{\mu^2(\Sm, ~T)}{\mu_0^2} = 0 ~.
\end{equation*}
Taking a time derivative of the above equation gives us
\begin{equation}\label{eq:yieldderiv}
  3~\Ss~\Ssdot - 2~\Sy~\cfrac{\mu^2}{\mu_0^2}~\Partial{\Sy}{t}
    - 2~\Sy^2~\cfrac{\mu}{\mu_0^2}~\Partial{\mu}{\Sm}~\Smdot = 0~.
\end{equation}

We assume that the plastic part of the rate of deformation can be
obtained from a plastic potential in the form (associated flow rule)
\[
  \Bd^p = \Lambdadot~\Partial{f}{\Bsig} = \Lambdadot~f_{\sigma}
\]
The unit outward normal to the yield surface is given by
\[
  \BnHat = \cfrac{1}{\xi}\Partial{f}{\Bsig} = \cfrac{1}{\xi}~f_{\sigma}~;~~ 
    \xi = \Norm{\Partial{f}{\Bsig}} = \Norm{f_{\sigma}}~;~~ \BnHat:\BnHat = 1
   \qquad \implies \qquad
   \Bd^p = \Lambdadot~\xi~\BnHat ~.
\]
Therefore,
\begin{equation*}
  \Epdot{p} := \sqrt{\cfrac{2}{3}}~ \Lambdadot~\xi
  ~;~~
  \Ep := \sqrt{\cfrac{2}{3}} \int_0^t \Lambdadot~\xi~dt^{'}  ~.
\end{equation*}
Note that
\[
  \Partial{\Sm}{\Bsig} = \BoneHat ~;~~ \Partial{\Ss}{\Bsig} = \BsHat ~.
\]
Assuming that $\Epdot{}$, $\Ep$, and $T$ remain unchanged during the 
elastic part of the deformation, we have
\[
  f_{\sigma} = 
  \Partial{f}{\Bsig} = \Partial{f}{\Sm}~\BoneHat + \Partial{f}{\Ss}~\BsHat
  = f_m~\BoneHat + f_s~\BsHat ~;~~
  \xi = \sqrt{(f_m)^2 + (f_s)^2} ~.
\]
For the yield function (\ref{eq:yieldFn}), we have
\[
  f_m =  - 2~\Sy^2(\Epdot,~\Ep,~T)~\cfrac{\mu(\Sm, ~T)}{\mu_0^2}~
                        \Partial{\mu}{\Sm} ~;~~
  f_s =  3~\Ss  ~.
\]
Therefore, equation (\ref{eq:yieldderiv}) can be written as
\begin{equation}\label{eq:yieldderiv1}
  f_s~\Ssdot - 2~\Sy~\cfrac{\mu^2}{\mu_0^2}~\Partial{\Sy}{t} + f_m~\Smdot = 0~.
\end{equation}
Also,
\[
  \Bd^p = \Lambdadot~
    \LBs - 2~\Sy^2(\Epdot,~\Ep,~T)~\cfrac{\mu(\Sm, ~T)}{\mu_0^2}~
         \Partial{\mu}{\Sm} \RBs \BoneHat + 
    3~\Lambdadot~\Ss~\BsHat ~. 
\]
This implies that
\begin{equation*}
  \Bd^p = (\Bd^p:\BoneHat)\BoneHat + (\Bd^p:\BsHat)\BsHat ~;~~
  \Bd^p:\BoneHat = - 2~\Lambdadot~
        \Sy^2(\Epdot,~\Ep,~T)~\cfrac{\mu(\Sm, ~T)}{\mu_0^2}~
         \Partial{\mu}{\Sm} ~;~~
  \Bd^p:\BsHat = 3~\Lambdadot~\Ss ~;~~
\end{equation*}
Recall, 
\begin{align*}
  \dot{\Bsig} & = \Smdot~\BoneHat + \Ssdot~\BsHat + \Ss~\dot{\BsHat} \\
  \Smdot & = 3~K(J^e)~\Bd^e:\BoneHat 
              - 2\Sthr~\rho_0~C_v~\Gamma_0~\dot{T} \\
  \Ssdot & = 2~\mu(\Sm, T)~\Bd^e:\BsHat \\
  \dot{\BsHat} & = \cfrac{2~\mu(\Sm,T)}{\Ss}~
                  \LBs \Bd^e - (\Bd^e:\BoneHat)\BoneHat
                       - (\Bd^e:\BsHat)\BsHat\RBs 
\end{align*}
Using the decomposition $\Bd^e = \Bd - \Bd^p$, we get
\begin{align*}
  \Smdot & = 3~K(J^e)~(\Bd - \Bd^p):\BoneHat
            - 2\Sthr~\rho_0~C_v~\Gamma_0~\dot{T} \\
  \Ssdot & = 2~\mu(\Sm, T)~(\Bd - \Bd^p):\BsHat \\
  \dot{\BsHat} & = \cfrac{2~\mu(\Sm,T)}{\Ss}~
                  \LBs \Bd - \Bd^p - (\Bd:\BoneHat)\BoneHat + 
                                     (\Bd^p:\BoneHat)\BoneHat
                  - (\Bd:\BsHat)\BsHat + (\Bd^p:\BsHat)\BsHat\RBs 
\end{align*}
or,
\[
  \begin{aligned}
  \dot{\Bsig} & = \Smdot~\BoneHat + \Ssdot~\BsHat + \Ss~\dot{\BsHat} \\
  \Smdot & = 3~K(J^e) \LBs~\Bd:\BoneHat  +
       2~\Lambdadot~ \Sy^2(\Epdot,~\Ep,~T)~\cfrac{\mu(\Sm, ~T)}{\mu_0^2}~
         \Partial{\mu}{\Sm} \RBs - 2\Sthr~\rho_0~C_v~\Gamma_0~\dot{T} \\
  \Ssdot & = 2~\mu(\Sm, T)~(\Bd:\BsHat - 3~\Lambdadot~\Ss) =
             2~\mu(\Sm, T)~(\Beta:\BsHat - 3~\Lambdadot~\Ss) \\
  \dot{\BsHat} & = \cfrac{2~\mu(\Sm,T)}{\Ss}~
                  \LBs \Beta - (\Bd:\BsHat)\BsHat \RBs 
                 = \cfrac{2~\mu(\Sm,T)}{\Ss}~
                   \LBs \Beta - (\Beta:\BsHat)\BsHat \RBs 
  \end{aligned}
\]
Now, the consistency condition requires that
\begin{equation*}
  \Partial{}{t}[f(\Sm,~\Ss,~\Epdot,~\Ep,~T)] = 0
\end{equation*}
or,
\begin{equation*}
  \Partial{f}{\Sm}~\Smdot + \Partial{f}{\Ss}~\Ssdot + 
    \Partial{f}{\Epdot{}}~\Partial{\Epdot{}}{t} + \Partial{f}{\Ep}~\Epdot{} +
    \Partial{f}{T}~\dot{T} = 0 
\end{equation*}
or,
\[
  f_m~\Smdot + f_s~\Ssdot + 
     f_d~\ddot{\epsilon} + f_p~\Epdot{} + f_t~\dot{T} = 0 
\]
where
\begin{align*}
  f_d & := \Partial{f}{\Epdot{}} = 2~\sigma_y~\cfrac{\mu^2}{\mu_0^2}~
     \Partial{\sigma_y}{\Epdot{}} ~; \quad
  \ddot{\epsilon}  = \sqrt{\cfrac{2}{3}}~ \cfrac{\dot{\Bd}:\Bd}{\Norm{\Bd}}~;~~
  \dot{\Bd} = \BR^T \cdot \Half[\Grad{\Ba} + (\Grad{\Ba})^T] \cdot \BR \\
  f_p & := \Partial{f}{\Ep} = 2~\sigma_y~\cfrac{\mu^2}{\mu_0^2}~
     \Partial{\sigma_y}{\Ep} ~; \qquad
  f_t := \Partial{f}{T} = 2~\sigma_y~\cfrac{\mu^2}{\mu_0^2}~
     \Partial{\sigma_y}{T} ~.
\end{align*}
Note that the above equation is the same as equation (\ref{eq:yieldderiv1}).
Also note that,
\begin{equation*}
  \Smdot = 3~K~[\Bd:\BoneHat - \Lambdadot~f_m] - G(T)~\dot{T} ~;~~
  \Ssdot = 2~\mu~[\Beta:\BsHat - \Lambdadot~f_s] 
\end{equation*}
where
\begin{equation*}
  G(T) := 2\Sthr~\rho_0~C_v(T)~\Gamma_0 ~.
\end{equation*}
Plugging into the consistency condition, we get
\begin{equation*}
  3~K~f_m~[\Bd:\BoneHat - \Lambdadot~f_m] - f_m~G(T)~\dot{T} +
  2~\mu~f_s~[\Beta:\BsHat - \Lambdadot~f_s] + 
     f_d~\ddot{\epsilon} + f_p~\Epdot{} + f_t~\dot{T} = 0 
\end{equation*}
or, 
\begin{equation*}
  \Lambdadot~(3~K~f_m^{~2} + 2~\mu~f_s^{~2}) = 
    3~K~f_m~\Bd:\BoneHat + 2~\mu~f_s~\Beta:\BsHat + 
     f_d~\ddot{\epsilon} + f_p~\Epdot{} + [f_t - f_m~G(T)]~\dot{T} = 0 
\end{equation*}
or,
\[
  \Lambdadot = \cfrac
    {3~K~f_m~\Bd:\BoneHat + 2~\mu~f_s~\Beta:\BsHat + 
     f_d~\ddot{\epsilon} + f_p~\Epdot{} + [f_t - f_m~G(T)]~\dot{T}}
    {3~K~f_m^{~2} + 2~\mu~f_s^{~2}} ~.
\]

\subsection{Mass Balance:}
The mass balance equation is
\[
  \dot{\rho} + \rho~\Div{\Bv} = 0 
\]
where $\rho$ is the current mass density, and $\dot{\rho}$ is the material
time derivative of $\rho$.

\subsection{Momentum Balance:}
The balance of linear momentum can be written as
\[
  \Div{\Bsig} + \rho~\Bb  = \rho~\dot{\Bv}
\]
where $\Bsig$ is the Cauchy stress, $\Bb$ is the body force per unit volume,
and $\dot{\Bv}$ is the acceleration.

The momentum equation can be written as
\begin{equation*}
  \Div{(p~\Bone + \Bs)} + \rho~\Bb  = \rho~\dot{\Bv}
\end{equation*}
or,
\[
  \Div{(p~\Bone)} + \Div{\Bs} + \rho~\Bb  = \rho~\dot{\Bv} ~.
\]

\subsection{Energy Balance:}
The balance of energy can be written as
\[
  \Bsig:\Bd - \Div{\Bq} + \rho~\dot{h} = \rho~\dot{e}
\]
where $\Bq$ is the heat flux per unit area,
$h$ is the heat source per unit mass, and $e$ is the specific internal energy.  
A dot over a quantity represents the material time derivative of that quantity.

We convert the energy equation into an approximate equation for heat 
conduction that includes heat source term due to plastic dissipation 
(\cite{Perzyna93}):
\[
  \rho~C_v~\dot{T} - \Div{(\Bdot{\Bkappa}{\Grad{T}})} - \zeta~\Bsig:\Bd^p = 
  \rho~\dot{h}
\]
where $C_v$ is the specific heat at constant volume, $\Bkappa$ is a 
second order tensor of thermal conductivity, and $\zeta$ is the Taylor-Quinney
coefficient.  Expressing the stress and the plastic part of rate of 
deformation into volumetric and deviatoric parts, we get
\begin{equation*}
  \rho~C_v~\dot{T} - \Div{(\Bdot{\Bkappa}{\Grad{T}})} - 
  \zeta~(p~\Bone + \Bs):\LBr\Third~\Tr{\Bd^p}~\Bone + \Beta^p\RBr = 
  \rho~\dot{h}
\end{equation*}
or, 
\begin{equation*}
  \rho~C_v~\dot{T} - \Div{(\Bdot{\Bkappa}{\Grad{T}})} - 
  \zeta~\LBr\Third~p~\Tr{\Bd^p}\Bone:\Bone + \Third~\Tr{\Bd^p}~\Bs:\Bone + 
        p~\Bone:\Beta^p + \Bs:\Beta^p\RBr = 
  \rho~\dot{h}
\end{equation*}
or, 
\[
  \rho~C_v~\dot{T} - \Div{(\Bdot{\Bkappa}{\Grad{T}})} - 
  \zeta~(p~\Tr{\Bd^p} + \Bs:\Beta^p) = \rho~\dot{h}
\]

\section{Weak Forms}
Let us now derive the weak forms of the governing equations.
\subsection{Rate of deformation:}
The weak form of the kinematic relation is given by
\[
  \IntOmega \BWs:\LBc\Bd - \Half\LBs\Gradv + \GradvT\RBs\RBc~d\Omega
     = 0
\]
where $\BWs$ is an arbitrary second-order tensor valued weighting function.  
We write the rate of deformation and the weighting function in terms of their
volumetric and deviatoric components to get
\begin{equation*}
  \IntOmega \LBc\BWsDev+\BWsVol\RBc:\LBc\Beta + 
      \Third~\ThetaDot~\Bone - 
      \Half\LBs\Gradv + \GradvT\RBs\RBc~d\Omega = 0
\end{equation*}
{\footnotesize
or
\begin{align*}
  \IntOmega & \LBc
    \BWsDev:\Beta + \Third~\ThetaDot~\BWsDev:\Bone - 
      \Half\BWsDev:\LBs\Gradv + \GradvT\RBs\RBn + \\
    & \LBn\BWsVol:\Beta + \Ninth~\ThetaDot~\Tr{\BWs}\Bone:\Bone - 
      \Sixth~\Tr{\BWs}\Bone:\LBs\Gradv + \GradvT\RBs\RBc~d\Omega = 0
\end{align*}
or
\begin{equation*}
  \IntOmega \LBc
    \BWsDev:\Beta - \Half\BWsDev:\LBs\Gradv + \GradvT\RBs + 
    \Third~\ThetaDot~\Tr{\BWs} - 
      \Sixth~\Tr{\BWs}\LBs\Tr{\Gradv} + \Tr{\GradvT}\RBs\RBc~d\Omega = 0
\end{equation*}
or 
\begin{equation*}
  \IntOmega \LBc
    \BWsDev:\LBr\Beta - \Bd\RBr + 
    \Third~\ThetaDot~\Tr{\BWs} - \Third~\Tr{\BWs}~\Div{\Bv}\RBc~d\Omega = 0
\end{equation*}
or 
\begin{equation*}
  \IntOmega \LBc
    -\Third~\ThetaDot~\BWsDev:\Bone + 
    \Third~\ThetaDot~\Tr{\BWs} - \Third~\Tr{\BWs}~\Div{\Bv}\RBc~d\Omega = 0
\end{equation*}
}
or
\begin{equation*}
  \IntOmega \LBc
    \Third~\ThetaDot~\Tr{\BWs} - \Third~\Tr{\BWs}~\Div{\Bv}\RBc~d\Omega = 0
\end{equation*}
Define $w^* := 1/3~\Tr{\BWs}$.  Then the weak form becomes
\[
  \IntOmega w^*\LBr\ThetaDot - \Div{\Bv}\RBr~d\Omega = 0
\]

\subsection{Constitutive Relations:}
The weak forms of the constitutive relations are given by
\[
  \IntOmega \BWss:\LBr\Truesdell{\Bs} - \SfC(p, T):\Beta^e\RBr
     ~d\Omega = 0 
\]
and
\[
  \IntOmega w^{**}\LBs p - 
     \frac{\rho_0 C_0^2 (1 - J) [1 - \Gamma_0 (1 - J)]}
           {[1 - S_{\alpha}(1 - J)]^2} - 2~\Gamma_0 e \RBs
     ~d\Omega = 0 
\]
where $\BWss$ is a tensor valued weighting function and $w^{**}$ is a 
scalar weighting function.

\subsection{Momentum Balance:}
To derive the weak form of the momentum equation, we multiply the 
momentum equation with a vector-valued weighting function ($\Bw$) and
integrate over the domain ($\Omega$).  The weighting function ($\Bw$)
satisfies velocity boundary conditions on the parts of the boundary
where velocities are prescribed.  Then we get,
\[
  \IntOmega \Bdot{\Bw}
                 {\LBs\Div{(p~\Bone)} + \Div{\Bs} + \rho~\Bb\RBs}~d\Omega 
     = \IntOmega \rho~\Bdot{\Bw}{\dot{\Bv}}~d\Omega ~.
\]
Using the identity 
$\Bdot{\Bv}{(\Div{\BS)}} = \Div{(\Bdot{\BS^T}{\Bv})} - \BS:\Grad{\Bv}$, 
where $\BS$ is a second-order tensor valued field and $\Bv$ is a 
vector valued field, we get (using the symmetry of the stress tensor)
\begin{equation*}
  \IntOmega \LBc
     \Div{[\Bdot{(p~\Bone)}{\Bw}]} - p~\Bone:\Grad{\Bw} +
     \Div{(\Bdot{\Bs}{\Bw})} - \Bs:\Grad{\Bw} +
     \rho~\Bdot{\Bw}{\Bb}\RBc~d\Omega 
     = \IntOmega \rho~\Bdot{\Bw}{\dot{\Bv}}~d\Omega
\end{equation*}
or,
\[
  \IntOmega \LBc
     \Div{(p~\Bw)} - p~\Div{\Bw} +
     \Div{(\Bdot{\Bs}{\Bw})} - \Bs:\Grad{\Bw} +
     \rho~\Bdot{\Bw}{\Bb}\RBc~d\Omega 
     = \IntOmega \rho~\Bdot{\Bw}{\dot{\Bv}}~d\Omega ~.
\]
%Using the identity 
%$\Div{(\phi~\Bw)} = \Bdot{\Bw}{\Grad{\phi}} + \phi~\Div{\Bw}$ where
%$\phi$ is a scalar valued field and $\Bw$ is a vector valued field, we get
%\begin{equation*}
%  \IntOmega \LBc
%     \Bdot{\Bw}{\Grad{p}} + p~\Div{\Bw} - p~\Div{\Bw} +
%     \Div{(\Bdot{\Bs}{\Bw})} - \Bs:\Grad{\Bw} +
%     \rho~\Bdot{\Bw}{\Bb}\RBc~d\Omega 
%     = \IntOmega \rho~\Bdot{\Bw}{\dot{\Bv}}~d\Omega 
%\end{equation*}
%or,
%\begin{equation}
%  \IntOmega \LBc
%     \Bdot{\Bw}{\Grad{p}} + 
%     \Div{(\Bdot{\Bs}{\Bw})} - \Bs:\Grad{\Bw} +
%     \rho~\Bdot{\Bw}{\Bb}\RBc~d\Omega 
%     = \IntOmega \rho~\Bdot{\Bw}{\dot{\Bv}}~d\Omega ~.
%\end{equation}
Applying the divergence theorem, we have 
\begin{equation*}
  \IntOmega \LBc
     - p~\Div{\Bw} - \Bs:\Grad{\Bw} +
     \rho~\Bdot{\Bw}{\Bb}\RBc~d\Omega +
  \IntGamma \Bdot{\Bn}{(p~\Bw)} ~d\Gamma + 
  \IntGamma \Bdot{\Bn}{(\Bdot{\Bs}{\Bw})} ~d\Gamma
     = \IntOmega \rho~\Bdot{\Bw}{\dot{\Bv}}~d\Omega 
\end{equation*}
{\footnotesize
or, 
\begin{equation*}
  \IntOmega \LBc\rho~\Bdot{\Bw}{\dot{\Bv}} 
     + p~\Div{\Bw} + \Bs:\Grad{\Bw}\RBc~d\Omega = 
  \IntOmega \rho~\Bdot{\Bw}{\Bb}~d\Omega +
  \IntGamma p~\Bdot{\Bn}{\Bw} ~d\Gamma + 
  \IntGamma \Bdot{\Bn}{(\Bdot{\Bs}{\Bw})} ~d\Gamma
\end{equation*}
or, 
\begin{equation*}
  \IntOmega \LBc\rho~\Bdot{\Bw}{\dot{\Bv}} 
     + p~\Div{\Bw} + \Bs:\Grad{\Bw}\RBc~d\Omega = 
  \IntOmega \rho~\Bdot{\Bw}{\Bb}~d\Omega +
  \IntGamma p~\Bdot{\Bn}{\Bw} ~d\Gamma + 
  \IntGamma \Bdot{(\Bdot{\Bs^T}{\Bn})}{\Bw} ~d\Gamma
\end{equation*}
or, 
\[
  \IntOmega \LBc\rho~\Bdot{\Bw}{\dot{\Bv}} 
     + p~\Div{\Bw} + \Bs:\Grad{\Bw}\RBc~d\Omega = 
  \IntOmega \rho~\Bdot{\Bw}{\Bb}~d\Omega +
  \IntGamma p~\Bdot{\Bn}{\Bw} ~d\Gamma + 
  \IntGamma \Bdot{(\Bdot{\Bsig^T}{\Bn} - p~\Bn)}{\Bw} ~d\Gamma 
\]
}
or,
\[
  \IntOmega \LBc\rho~\Bdot{\Bw}{\dot{\Bv}} 
     + p~\Div{\Bw} + \Bs:\Grad{\Bw}\RBc~d\Omega = 
  \IntOmega \rho~\Bdot{\Bw}{\Bb}~d\Omega +
  \IntGamma \Bdot{(\Bdot{\Bsig^T}{\Bn})}{\Bw} ~d\Gamma ~.
\]
In the above, $\Bn$ is the outward normal to the surface $\Gamma$.

If the applied surface traction is $\Bart = \Bdot{\Bsig^T}{\Bn}$, we
get (since $\Bw$ is zero on the part of the boundary where velocities
are specified)
\[
  \IntOmega \LBc\rho~\Bdot{\Bw}{\dot{\Bv}} 
     + p~\Div{\Bw} + \Bs:\Grad{\Bw}\RBc~d\Omega = 
  \IntOmega \rho~\Bdot{\Bw}{\Bb}~d\Omega +
  \IntGammat \Bdot{\Bart}{\Bw}~d\Gamma ~.
\]

\subsection{Energy Balance:}
The weak form of the heat conduction equation is given by
\begin{equation*}
  \IntOmega \What\LBs\rho~C_v~\dot{T} - \Div{(\Bdot{\Bkappa}{\Grad{T}})} - 
  \zeta~(p~\Tr{\Bd^p} + \Bs:\Beta^p)\RBs~d\Omega = 
  \IntOmega \What~\rho~\dot{h}~d\Omega~.
\end{equation*}
{\footnotesize
Using the identity $\phi (\Div{\Bv}) = \Div{(\phi~\Bv)} - 
\Bdot{(\Grad{\phi})}{\Bv}$ we get
\begin{equation*}
  \IntOmega \What~\rho~C_v~\dot{T}~d\Omega - 
  \IntOmega \LBc\Div{[\What(\Bdot{\Bkappa}{\Grad{T}})]} - 
            \Bdot{(\Grad{\What})}{\Bdot{\Bkappa}{\Grad{T}}}\RBc~d\Omega -
  \IntOmega \What~\zeta~[p~\Tr{\Bd^p} + \Bs:\Beta^p]~d\Omega = 
  \IntOmega \What~\rho~\dot{h}~d\Omega~.
\end{equation*}
Using the divergence theorem, we have
\begin{equation*}
  \IntOmega \LBc\What~\rho~C_v~\dot{T} + 
            \Bdot{(\Grad{\What})}{\Bdot{\Bkappa}{\Grad{T}}}\RBc~d\Omega =
  \IntOmega \What~\LBc\rho~\dot{h} + 
              \zeta~[p~\Tr{\Bd^p} + \Bs:\Beta^p]\RBc~d\Omega  + 
  \IntGamma \Bn\cdot[\What(\Bdot{\Bkappa}{\Grad{T}})]~d\Gamma ~.
\end{equation*}
}
If a heat flux ($\Barq = \Bn\cdot(\Bkappa\cdot\Grad{T})$) is specified 
on part of the boundary ($\Gamma_q$), the weak form becomes
\[
  \IntOmega \LBc\What~\rho~C_v~\dot{T} + 
            \Bdot{(\Grad{\What})}{\Bdot{\Bkappa}{\Grad{T}}}\RBc~d\Omega =
  \IntOmega \What~\LBc\rho~\dot{h} + 
              \zeta~[p~\Tr{\Bd^p} + \Bs:\Beta^p]\RBc~d\Omega  + 
  \IntGammaq \What~\Barq~d\Gamma ~.
\]

\section{MPM Discretization}
We can now discretize each of the weak forms using appropriate basis functions.
\subsection{Rate of Deformation:}
The weak form of the rate of deformation equation is
\[
  \IntOmega w^*\LBr\ThetaDot - \Div{\Bv}\RBr~d\Omega = 0 ~.
\]
We assume a discontinuous approximation for $\ThetaDot$ over
each cell and a weighting function $w^*$ of the form \footnote{ 
The approximations for $w^*$ and $\ThetaDot$ need not
be a sum over the particles.  We could alternatively assume a constant
value over each cell or a sum over the grid nodes with the same basis
functions as the grid basis functions.  We have chosen a sum over particles
in a cell to keep our formulation consistent with the standard MPM 
procedure.}
\[
  \ThetaDot(\Bx) \approx \Sumpcq \ThetaDot_q \chi_q(\Bx) ~;\qquad
  w^*(\Bx) = \Sumpc w^*_p \chi_p(\Bx) 
\]
where $n_p^c$ is the number of particles in a grid cell, $w^*_p$ are the
weighting functions at the particle locations, and $\chi_p(\Bx)$ are the 
basis functions.  The velocities are approximated in the standard MPM way using
\[
  \Bv(\Bx) \approx \Sumg \Bv_g N_g(\Bx) 
\]
where $n_g$ is the number of grid points, $\Bv_g$ are the velocities at the
grid vertices, and $N_g(\Bx)$ are the grid basis functions.

Plugging the approximations into the weak form, we get the following equations
that are valid over each cell
{\footnotesize
\begin{equation*}
  \IntOmegac \LBs\Sumpc w^*_p \chi_p(\Bx)\RBs
    \LBc\Sumpcq \ThetaDot_q \chi_q(\Bx) - 
    \Div{\LBs\Sumgc \Bv_g N_g(\Bx) \RBs}\RBc~d\Omega = 0 
\end{equation*}
or,
\begin{equation*}
  \Sumpc w^*_p \LBs\IntOmegac \chi_p(\Bx)
    \LBc\Sumpcq \ThetaDot_q \chi_q(\Bx) - 
    \Sumgc \Bdot{\Bv_g}{\Grad{N_g}}\RBc~d\Omega\RBs = 0  ~.
\end{equation*}
The abitrariness of $w^*_p$ gives us a system of $n_p^c$ equations
\begin{equation*}
  \IntOmegac \chi_p(\Bx)
    \LBc\Sumpcq \ThetaDot_q \chi_q(\Bx) - 
    \Sumgc \Bdot{\Bv_g}{\Grad{N_g}}\RBc~d\Omega = 0 
    ~;\qquad p = 1\dots n_p^c
\end{equation*}
or,
\begin{equation*}
  \Sumpcq \LBs\IntOmegac \chi_p(\Bx) \chi_q(\Bx)~d\Omega\RBs \ThetaDot_q
   - \Sumgc \LBs\IntOmegac \chi_p(\Bx) \Grad{N_g}~d\Omega\RBs\cdot\Bv_g = 0
   ~; \qquad p = 1\dots n_p^c
\end{equation*}
or, 
}
\begin{equation*}
  \Sumpcq \LBs\IntOmegac \chi_p(\Bx) \chi_q(\Bx)~d\Omega\RBs \ThetaDot_q
   = \Sumgc \LBs\IntOmegac \chi_p(\Bx) \Grad{N_g}~d\Omega\RBs\cdot\Bv_g  ~;
   \qquad p = 1\dots n_p^c ~.
\end{equation*}
If we identify the basis functions $\chi_p$ with the particle characteristic
functions and use the property (\cite{Bard04}, p. 485)
\[
  \chi_p(\Bx) = \begin{cases}
                  1 & \text{if}~\Bx \in \Omega_p \\
                  0 & \text{otherwise}
                \end{cases}
\]
we get
\begin{equation*}
  \LBs\IntOmegapc \chi_p(\Bx) \chi_p(\Bx)~d\Omega\RBs \ThetaDot_p
   = \Sumgc \LBs\IntOmegapc \chi_p(\Bx) \Grad{N_g}~d\Omega\RBs
      \cdot\Bv_g  ~;
   \qquad p = 1\dots n_p^c ~.
\end{equation*}
Define the volume of the particle inside the cell as
\[
  V_{pc} := \IntOmegapc \chi_p(\Bx) \chi_p(\Bx)~d\Omega ~.
\]
We then have
\begin{equation*}
  V_{pc}~\ThetaDot_p
   = \Sumgc \LBs\IntOmegapc \chi_p(\Bx) \Grad{N_g}~d\Omega\RBs
      \cdot\Bv_g  ~;
   \qquad p = 1\dots n_p^c ~.
\end{equation*}
Define the gradient weighting function as
\begin{equation}\label{eq:gradS}
   \Grad{N_{pg}} := \IntOmegapc \chi_p(\Bx) \Grad{N_g}~d\Omega ~.
\end{equation}
Then,
\begin{equation*}
  V_{pc}~\ThetaDot_p
   = \Sumgc \Grad{N_{pg}}\cdot\Bv_g  ~;
   \qquad p = 1\dots n_p^c ~.
\end{equation*}
or,
\[
  \ThetaDot_p = \cfrac{1}{V_{pc}}\Sumgc \Grad{N_{pg}}\cdot\Bv_g~;
   \qquad p = 1\dots n_p^c 
\]

\subsection{Constitutive Relations:}
The weak form of the constitutive relation for pressure is
\[
  \IntOmega w^{**}(\Bx)\LBs p(\Bx) - 
     \frac{\rho_0 C_0^2 (1 - J(\Bx)) [1 - \Gamma_0 (1 - J(\Bx))]}
           {[1 - S_{\alpha}(1 - J(\Bx))]^2} - 2~\Gamma_0 e(\Bx) \RBs
     ~d\Omega = 0 ~.
\]
We use a discontinuous approximation for the pressure field ($p$) and
a weighting function $w^{**}$ of the form
\[
  p(\Bx) \approx \Sumpcq p_q \chi_q(\Bx) ~;\qquad
  w^{**}(\Bx) = \Sumpc w^{**}_p \chi_p(\Bx) ~.
\]
After plugging these approximations into the weak form, we get the 
following relations over each cell
{\footnotesize
\begin{equation*}
  \IntOmegac \LBs\Sumpc w^{**}_p \chi_p(\Bx)\RBs
     \LBc \Sumpcq p_q \chi_q(\Bx) -
     \frac{\rho_0 C_0^2 (1 - J(\Bx)) [1 - \Gamma_0 (1 - J(\Bx))]}
           {[1 - S_{\alpha}(1 - J(\Bx))]^2} - 2~\Gamma_0 e(\Bx) \RBc
     ~d\Omega = 0
\end{equation*}
or,
\begin{equation*}
  \Sumpc w^{**}_p \LBs\IntOmegac \chi_p(\Bx)
     \LBc \Sumpcq p_q \chi_q(\Bx) -
     \frac{\rho_0 C_0^2 (1 - J(\Bx)) [1 - \Gamma_0 (1 - J(\Bx))]}
           {[1 - S_{\alpha}(1 - J(\Bx))]^2} - 2~\Gamma_0 e(\Bx) \RBc
     ~d\Omega\RBs = 0 ~.
\end{equation*}
Invoking the arbitrariness of $w^{**}_p$ we get
\begin{equation*}
  \IntOmegac \chi_p(\Bx)
     \LBc \Sumpcq p_q \chi_q(\Bx) -
     \frac{\rho_0 C_0^2 (1 - J(\Bx)) [1 - \Gamma_0 (1 - J(\Bx))]}
           {[1 - S_{\alpha}(1 - J(\Bx))]^2} - 2~\Gamma_0 e(\Bx) \RBc
     ~d\Omega = 0 ~;~~ p=1\dots n_p^c
\end{equation*}
or, 
\begin{equation*}
  \Sumpcq \LBs\IntOmegac \chi_p(\Bx)~\chi_q(\Bx)~d\Omega\RBs~p_q 
    = \IntOmegac \chi_p(\Bx)\LBs
     \frac{\rho_0 C_0^2 (1 - J(\Bx)) [1 - \Gamma_0 (1 - J(\Bx))]}
           {[1 - S_{\alpha}(1 - J(\Bx))]^2} - 2~\Gamma_0 e(\Bx)\RBs
     ~d\Omega  ~;~~ p=1\dots n_p^c
\end{equation*}
}
Once again, if we identify the basis functions $\chi_p$ with the particle 
characteristic functions and use the property
\[
  \chi_p(\Bx) = \begin{cases}
                  1 & \text{if}~\Bx \in \Omega_p \\
                  0 & \text{otherwise}
                \end{cases}
\]
and we get
{\footnotesize
\begin{equation*}
  \begin{aligned}
  \LBs\IntOmegapc \chi_p(\Bx)~\chi_p(\Bx)~d\Omega\RBs~p_p 
    & = \IntOmegapc \chi_p(\Bx)\LBs
     \frac{\rho_0 C_0^2 (1 - J(\Bx)) [1 - \Gamma_0 (1 - J(\Bx))]}
           {[1 - S_{\alpha}(1 - J(\Bx))]^2} - 2~\Gamma_0 e(\Bx)\RBs\,d\Omega \\
    & \qquad \qquad   p=1\dots n_p^c 
  \end{aligned}
\end{equation*}
or,
\begin{equation*}
   V_{pc}~p_p 
    = \IntOmegapc \chi_p(\Bx)\LBs
     \frac{\rho_0 C_0^2 (1 - J(\Bx)) [1 - \Gamma_0 (1 - J(\Bx))]}
           {[1 - S_{\alpha}(1 - J(\Bx))]^2} - 2~\Gamma_0 e(\Bx)\RBs
     ~d\Omega  ~;~~ p=1\dots n_p^c ~.
\end{equation*}
}
Identifying the volume averaged $J(\Bx)$ and $e(\Bx)$ as $\theta_p$ and
$e_p$, respectively, we get
\begin{equation*}
  V_{pc}~p_p 
    = V_{pc} \frac{\rho_0 C_0^2 (1 - \theta_p) [1 - \Gamma_0 (1 - \theta_p)]}
           {[1 - S_{\alpha}(1 - \theta_p)]^2} - 2~\Gamma_0 e_p
      ~;~~ p=1\dots n_p^c 
\end{equation*}
or, 
\[
  p_p = \frac{\rho_0 C_0^2 (1 - \theta_p) [1 - \Gamma_0 (1 - \theta_p)]}
           {[1 - S_{\alpha}(1 - \theta_p)]^2} - 2~\Gamma_0 e_p
      ~;~~ p=1\dots n_p^c ~.
\]
The weak form of the constitutive relation for the deviatoric stress rate
is given by
\[
  \IntOmega \BWss(\Bx):\LBs\Truesdell{\Bs}(\Bx) - 
      \SfC(p, T, \Bx):\Beta^e(\Bx)\RBs ~d\Omega = 0 ~.
\]
We assume that the weighting function is
\[
  \BWss(\Bx) = \Sumpq \BWssq \chi_q(\Bx)
\]
and plug these into the weak form to get the following relations 
\begin{equation*}
  \IntOmega \LBs\Sumpq \BWssq~\chi_q(\Bx)\RBs:
     \LBs\Truesdell{\Bs}(\Bx) - \SfC(p, T, \Bx):\Beta^e(\Bx)\RBs 
     ~d\Omega = 0
\end{equation*}
or, 
\begin{equation*}
  \Sumpq \BWssq : \LBs\IntOmega \chi_q(\Bx)
     \LBs\Truesdell{\Bs}(\Bx) - \SfC(p, T, \Bx):\Beta^e(\Bx)\RBs 
     ~d\Omega\RBs = 0 ~.
\end{equation*}
The arbitrariness of $\BWssq$ gives 
\begin{equation*}
  \IntOmegaq \chi_q(\Bx)
     \LBs\Truesdell{\Bs}(\Bx) - \SfC(p, T, \Bx):\Beta^e(\Bx)\RBs 
     ~d\Omega = 0 ~; \qquad q = 1\dots n_p
\end{equation*}
Using arguments similar to those used for the pressure constitutive relation,
we get
\[
  \Truesdell{\Bs}_q = \SfC_q(p, T):\Beta^e_q 
     ~; \qquad q = 1\dots n_p
\]

\subsection{Momentum Balance:}
The weak form of the momentum balance relation is
\[
  \IntOmega \LBs\rho(\Bx)~\Bdot{\Bw(\Bx)}{\Bvdot(\Bx)} 
     + p(\Bx)~\Div{\Bw} + \Bs(\Bx):\Grad{\Bw}\RBs~d\Omega = 
  \IntOmega \rho(\Bx)~\Bdot{\Bw(\Bx)}{\Bb(\Bx)}~d\Omega +
  \IntGammat \Bdot{\Bart(\Bx)}{\Bw(\Bx)}~d\Gamma ~.
\]
The first step in the MPM discretization is to convert the integral 
over $\Omega$ into a sum of integrals over particles.  To achieve this
we assume that the particle characteristic functions have the form
\[
  \chi_p(\Bx) = \begin{cases}
                  1 & \text{if}~\Bx \in \Omega_p \\
                  0 & \text{otherwise}
                \end{cases}
\]
Then the weak form of the momentum balance equation can be written as
\begin{align*}
  \Sump \IntOmegap & \chi_p(\Bx)\LBs\rho(\Bx)~\Bdot{\Bw(\Bx)}{\Bvdot(\Bx)} 
     + p(\Bx)~\Div{\Bw} + \Bs(\Bx):\Grad{\Bw}\RBs~d\Omega = \\
   & \Sump \IntOmegap \chi_p(\Bx)~\rho(\Bx)~\Bdot{\Bw(\Bx)}{\Bb(\Bx)}~d\Omega +
  \IntGammat \Bdot{\Bart(\Bx)}{\Bw(\Bx)}~d\Gamma ~.
\end{align*}
The weighting function is 
\[
  \Bw(\Bx) = \Sumg \Bw_g N_g(\Bx) ~.
\]
The velocity is approximated as
\[
  \Bv(\Bx) \approx \Sumgh \Bv_h N_h(\Bx) ~.
\]
The material time derivative of $\Bv$ is also approximated in a similar
manner (see \cite{Sulsky95}).  Thus,
\[
  \Bvdot(\Bx) \approx \Sumgh \Bvdot_h N_h(\Bx) ~.
\]
Plugging these into the left hand side of the momentum equation we get
{\footnotesize
\begin{align*}
  \text{LHS} = 
   \Sump \IntOmegap \chi_p(\Bx) & \LBs\rho(\Bx)~
     \LBc\Sumg \Bw_g N_g(\Bx)\RBc\cdot
     \LBc\Sumgh \Bvdot_h N_h(\Bx)\RBc  
     + p(\Bx)~\Div{\LBc\Sumg \Bw_g N_g(\Bx)\RBc}\RBn \\
     & \LBn+ \Bs(\Bx):\Grad{\LBc\Sumg \Bw_g N_g(\Bx)\RBc}
     \RBs~d\Omega 
\end{align*}
or, 
\begin{align*}
  \text{LHS} = 
    \Sumg & \Bw_g \cdot \LBs\Sump 
      \LBc\Sumgh
        \LBr\IntOmegap \chi_p(\Bx)~\rho(\Bx)~N_g(\Bx)~N_h(\Bx)~d\Omega\RBr
        \Bvdot_h\RBc\RBs + \\
    & \Sumg \Bw_g \cdot \LBs\Sump
      \IntOmegap \chi_p(\Bx)~p(\Bx)~\Grad{N_g}~d\Omega \RBs + 
    \Sumg \Bw_g \cdot \LBs\Sump
      \IntOmegap \chi_p(\Bx)~\Bs(\Bx)\cdot\Grad{N_g}~d\Omega \RBs
\end{align*}
or, 
\begin{align*}
  \text{LHS} = 
    \Sumg & \Bw_g \cdot \LBs\Sump \LBs \LBc
      \Sumgh
        \LBr\IntOmegap \chi_p(\Bx)~\rho(\Bx)~N_g(\Bx)~N_h(\Bx)~d\Omega\RBr
        \Bvdot_h
     \RBc + \RBn\RBn \\
     & \LBn\LBn\IntOmegap \chi_p(\Bx)~p(\Bx)~\Grad{N_g}~d\Omega + 
     \IntOmegap \chi_p(\Bx)~\Bs(\Bx)\cdot\Grad{N_g}~d\Omega \RBs\RBs
\end{align*}
Similarly, the right hand side of the weak form of momentum balance can be 
written as
\begin{equation*}
  \text{RHS} = 
  \Sump \IntOmegap \chi_p(\Bx)~\rho(\Bx)~
    \LBs\Sumg \Bw_g N_g(\Bx)\RBs\cdot~\Bb(\Bx)~d\Omega +
  \IntGammat \Bart(\Bx)\cdot \LBs\Sumg \Bw_g N_g(\Bx)\RBs~d\Gamma
\end{equation*}
or, 
\begin{equation*}
  \text{RHS} = 
  \Sumg \Bw_g \cdot \LBs\Sump 
     \IntOmegap \chi_p(\Bx)~\rho(\Bx)~N_g(\Bx)~\Bb(\Bx)~d\Omega\RBs +
  \Sumg \Bw_g \cdot \LBs\IntGammat \Bart(\Bx)~N_g(\Bx)~d\Gamma\RBs
\end{equation*}
or, 
\begin{equation*}
  \text{RHS} = 
  \Sumg \Bw_g \cdot \LBs\LBc\Sump 
     \IntOmegap \chi_p(\Bx)~\rho(\Bx)~N_g(\Bx)~\Bb(\Bx)~d\Omega\RBc +
     \IntGammat \Bart(\Bx)~N_g(\Bx)~d\Gamma\RBs
\end{equation*}
Combining the left and right hand sides and invoking the arbitrariness of
$\Bw_g$, we get
\begin{align*}
  \Sump  & \LBs \LBc
      \Sumgh
        \LBr\IntOmegap \chi_p(\Bx)~\rho(\Bx)~N_g(\Bx)~N_h(\Bx)~d\Omega\RBr
        \Bvdot_h
     \RBc + \RBn \\
     & \LBn\IntOmegap \chi_p(\Bx)~p(\Bx)~\Grad{N_g}~d\Omega + 
     \IntOmegap \chi_p(\Bx)~\Bs(\Bx)\cdot\Grad{N_g}~d\Omega \RBs = \\
   & \LBc\Sump 
     \IntOmegap \chi_p(\Bx)~\rho(\Bx)~N_g(\Bx)~\Bb(\Bx)~d\Omega\RBc +
     \IntGammat \Bart(\Bx)~N_g(\Bx)~d\Gamma ~;~~ g = 1 \dots n_g
\end{align*}
}
If $\rho(\Bx)$, $p(\Bx)$, $\Bs(\Bx)$, and $\Bb(\Bx)$ are assumed to be 
constant over each particle, we can replace these quantities with the values at 
the particles to get
\begin{align*}
  \Sump  & \LBs \LBc
      \Sumgh
        \LBr\rho_p\IntOmegap \chi_p(\Bx)~N_g(\Bx)~N_h(\Bx)~d\Omega\RBr
        \Bvdot_h
     \RBc + \RBn \\
     & \LBn p_p\IntOmegap \chi_p(\Bx)~\Grad{N_g}~d\Omega + 
       \Bs_p\cdot\IntOmegap \chi_p(\Bx)\Grad{N_g}~d\Omega \RBs = \\
   & \LBc\Sump 
       \rho_p~\Bb_p\IntOmegap \chi_p(\Bx)~N_g(\Bx)~d\Omega\RBc +
     \IntGammat \Bart(\Bx)~N_g(\Bx)~d\Gamma ~;~~ g = 1 \dots n_g ~.
\end{align*}
Then, using the definition of the gradient weighting function (\ref{eq:gradS}),
and defining 
\begin{equation}\label{eq:Spg}
   N_{pg} := \IntOmegap \chi_p(\Bx)~N_g(\Bx)~d\Omega 
\end{equation}
we get
\begin{align*}
  \Sumgh & \LBc\Sump
        \LBr\rho_p\IntOmegap \chi_p(\Bx)~N_g(\Bx)~N_h(\Bx)~d\Omega\RBr
         \RBc
        \Bvdot_h 
        + \LBc\Sump p_p \Grad{N_{pg}}\RBc + 
          \LBc\Sump \Bs_p \cdot \Grad{N_{pg}}\RBc = \\
   & \LBc\Sump 
       \rho_p~\Bb_p~N_{pg}\RBc +
     \IntGammat \Bart(\Bx)~N_g(\Bx)~d\Gamma ~;~~ g = 1 \dots n_g
\end{align*}
Define the mass matrix as
\[
  m_{gh} := \Sump \rho_p\IntOmegap \chi_p(\Bx)~N_g(\Bx)~N_h(\Bx)~d\Omega ~.
\]
Define the internal force vector as
\[
  \Bf_g^{\Tint} := \Sump \LBs p_p \Grad{N_{pg}} + 
                             \Bs_p \cdot \Grad{N_{pg}}\RBs ~.
\]
Define the body force vector as
\[
  \Bf_g^{\Tbod} := \Sump \rho_p~\Bb_p~N_{pg} ~.
\]
Also define the external force vector as
\[
  \Bf_g^{\Text} := \IntGammat \Bart(\Bx)~N_g(\Bx)~d\Gamma 
\]
Then we get the semidiscrete system of equations
\[
  \Sumgh m_{gh} \Bvdot_h + \Bf_g^{\Tint} = \Bf_g^{\Tbod} + \Bf_g^{\Text}
      ~;~~ g = 1 \dots n_g
\]

\subsection{Energy Balance:}
The weak form of the energy balance equation is
\begin{align}
  \IntOmega & \LBc\What(\Bx)~\rho(\Bx)~C_v(\Bx)~\dot{T}(\Bx) + 
    \Bdot{\Grad{\What}}{\Bdot{\Bkappa(\Bx)}{\Grad{T}}}\RBc~d\Omega = \\
  & \IntOmega \What(\Bx)~\LBc\rho(\Bx)~\dot{h}(\Bx) + 
    \zeta~[p(\Bx)~\Tr{\Bd^p(\Bx)} + \Bs(\Bx):\Beta^p(\Bx)]\RBc ~d\Omega  + 
  \IntGammaq \What(\Bx)~\Barq(\Bx)~d\Gamma \nonumber~.
\end{align}
{\footnotesize
As before, we express this equation as a sum over particles to get
\begin{align*}
  \Sump \IntOmegap & \chi_p(\Bx)
      \LBc\What(\Bx)~\rho(\Bx)~C_v(\Bx)~\dot{T}(\Bx) + 
    \Bdot{\Grad{\What}}{\Bdot{\Bkappa(\Bx)}{\Grad{T}}}\RBc~d\Omega = \\
  & \Sump \IntOmegap \chi_p(\Bx)~\What(\Bx)~\LBc\rho(\Bx)~\dot{h}(\Bx) + 
    \zeta~[p(\Bx)~\Tr{\Bd^p(\Bx)} + \Bs(\Bx):\Beta^p(\Bx)]\RBc ~d\Omega  + 
  \IntGammaq \What(\Bx)~\Barq(\Bx)~d\Gamma 
\end{align*}
}
We approximate $T$ using
\[
   T(\Bx) \approx \Sumg T_g~N_g(\Bx) ~.
\]
The material time derivative of $T$ is then given by
\[
   \Tdot(\Bx) \approx \Sumg \Tdot_g~N_g(\Bx) ~.
\]
The weighting functions are also chosen to be of the same form:
\[
  \What(\Bx) = \Sumg \What_g~N_g(\Bx) ~.
\]
Plugging these into the weak form, we get
{\footnotesize
\begin{align*}
  \Sump \IntOmegap \chi_p(\Bx) & 
      \LBr\Sumg \What_g~N_g(\Bx)\RBr~\rho(\Bx)~C_v(\Bx)~
             \LBr\Sumgh \Tdot_h~N_h(\Bx)\RBr~d\Omega \\
  &  + \Sump \IntOmegap \chi_p(\Bx)~
     \Bdot{\Grad{\LBr\Sumg \What_g~N_g(\Bx)\RBr}}{\Bdot{\Bkappa(\Bx)}
      {\Grad{\LBr\Sumgh T_h~N_h(\Bx) \RBr}}}~d\Omega  \\
  = & \Sump \IntOmegap \chi_p(\Bx)\LBr\Sumg \What_g~N_g(\Bx)\RBr~\LBc
        \rho(\Bx)~\dot{h}(\Bx) + 
    \zeta~[p(\Bx)~\Tr{\Bd^p(\Bx)} + \Bs(\Bx):\Beta^p(\Bx)]\RBc ~d\Omega \\
  &  + \IntGammaq \LBr \Sumg \What_g~N_g(\Bx) \RBr ~\Barq(\Bx)~d\Gamma 
\end{align*}
For simplicity, let us assume that $\Bkappa(\Bx)$ is isotropic, i.e.,
$\Bkappa(\Bx) = \kappa(\Bx)$.  Then, 
\begin{align*}
  \Sumg \What_g & \LBc \Sump \IntOmegap \chi_p(\Bx) ~
      N_g(\Bx)~\rho(\Bx)~C_v(\Bx)~\LBr\Sumgh \Tdot_h~N_h(\Bx)\RBr~d\Omega\RBc \\
  + \Sumg \What_g & \LBc \Sump \IntOmegap \chi_p(\Bx)~\kappa(\Bx)~
      \Grad{N_g} \cdot \LBr \Sumgh \Tdot_h~\Grad{N_h}\RBr~d\Omega \RBc  \\
  = \Sumg \What_g & \LBs \Sump \IntOmegap \chi_p(\Bx)~N_g(\Bx)~\LBc
        \rho(\Bx)~\dot{h}(\Bx) + \zeta~[p(\Bx)~\Tr{\Bd^p(\Bx)} + 
        \Bs(\Bx):\Beta^p(\Bx)]\RBc ~d\Omega \RBs \\
  + \Sumg \What_g & \LBc \IntGammaq N_g(\Bx)~\Barq(\Bx)~d\Gamma \RBc
\end{align*}
Invoking the arbitrariness of $\What_g$, we get
\begin{align*}
  \Sump & \IntOmegap \chi_p(\Bx)~ 
      N_g(\Bx)~\rho(\Bx)~C_v(\Bx)~\LBr\Sumgh \Tdot_h~N_h(\Bx)\RBr~d\Omega 
  + \Sump \IntOmegap \chi_p(\Bx)~\kappa(\Bx)~
      \Grad{N_g} \cdot \LBr \Sumgh T_h~\Grad{N_h}\RBr~d\Omega  \\
  & = \Sump \IntOmegap \chi_p(\Bx)~N_g(\Bx)~\LBc
        \rho(\Bx)~\dot{h}(\Bx) + \zeta~[p(\Bx)~\Tr{\Bd^p(\Bx)} + 
        \Bs(\Bx):\Beta^p(\Bx)]\RBc ~d\Omega \\
  & + \IntGammaq N_g(\Bx)~\Barq(\Bx)~d\Gamma ~;~~ g = 1 \dots n_g
\end{align*}
or, 
\begin{align*}
  \Sumgh & \LBs \Sump 
     \IntOmegap \chi_p(\Bx)~N_g(\Bx)~N_h(\Bx)~\rho(\Bx)~C_v(\Bx)~d\Omega
     \RBs \Tdot_h
  + \Sumgh \LBs \Sump 
      \IntOmegap \chi_p(\Bx)~\kappa(\Bx)~\Grad{N_g}\cdot\Grad{N_h}~d\Omega
      \RBs T_h  \\
  & = \Sump \IntOmegap \chi_p(\Bx)~N_g(\Bx)~\LBc
        \rho(\Bx)~\dot{h}(\Bx) + \zeta~[p(\Bx)~\Tr{\Bd^p(\Bx)} + 
        \Bs(\Bx):\Beta^p(\Bx)]\RBc ~d\Omega \\
  & + \IntGammaq N_g(\Bx)~\Barq(\Bx)~d\Gamma ~;~~ g = 1 \dots n_g
\end{align*}
If we assume that $\rho(\Bx)$, $C_v(\Bx)$, $\kappa(\Bx)$, $\dot{h}(\Bx)$,
$p(\Bx)$, $\Bs(\Bx)$, and $\Bd^p(\Bx)$ are constant over each particle, we get
\begin{align*}
  \Sumgh & \LBs \Sump \rho_p~C_{vp}
     \IntOmegap \chi_p(\Bx)~N_g(\Bx)~N_h(\Bx)~d\Omega \RBs \Tdot_h
  + \Sumgh \LBs \Sump \kappa_p
      \IntOmegap \chi_p(\Bx)~\Grad{N_g}\cdot\Grad{N_h}~d\Omega \RBs T_h  \\
  & = \Sump \rho_p~\dot{h}_p \IntOmegap \chi_p(\Bx)~N_g(\Bx)~d\Omega + 
      \Sump \zeta~p_p~\Tr{\Bd^p_p} \IntOmegap \chi_p(\Bx)~N_g(\Bx)~d\Omega \\
  & + \Sump \zeta~\Bs_p:\Beta^p_p \IntOmegap \chi_p(\Bx)~N_g(\Bx) ~d\Omega 
    + \IntGammaq N_g(\Bx)~\Barq(\Bx)~d\Gamma ~;~~ g = 1 \dots n_g
\end{align*}
Using the definition of the particle weighting functions (\ref{eq:Spg}), we
have
\begin{align*}
  \Sumgh & \LBs \Sump \rho_p~C_{vp}
     \IntOmegap \chi_p(\Bx)~N_g(\Bx)~N_h(\Bx)~d\Omega \RBs \Tdot_h
  + \Sumgh \LBs \Sump \kappa_p
      \IntOmegap \chi_p(\Bx)~\Grad{N_g}\cdot\Grad{N_h}~d\Omega \RBs T_h  \\
  & = \Sump \rho_p~\dot{h}_p~N_{pg} + 
      \Sump \zeta~p_p~\Tr{\Bd^p_p}~N_{pg}
    + \Sump \zeta~\Bs_p:\Beta^p_p~N_{pg}
    + \IntGammaq N_g(\Bx)~\Barq(\Bx)~d\Gamma ~;~~ g = 1 \dots n_g
\end{align*}
}
or, 
\begin{align*}
  \Sumgh & \LBs \Sump \rho_p~C_{vp}
     \IntOmegap \chi_p(\Bx)~N_g(\Bx)~N_h(\Bx)~d\Omega \RBs \Tdot_h
  + \Sumgh \LBs \Sump \kappa_p
      \IntOmegap \chi_p(\Bx)~\Grad{N_g}\cdot\Grad{N_h}~d\Omega \RBs T_h  \\
  & = \Sump \LBs \rho_p~\dot{h}_p + \zeta~p_p~\Tr{\Bd^p_p}
        + \zeta~\Bs_p:\Beta^p_p \RBs N_{pg}
    + \IntGammaq N_g(\Bx)~\Barq(\Bx)~d\Gamma ~;~~ g = 1 \dots n_g ~.
\end{align*}
Define the thermal inertia matrix as
\[
   M_{gh} :=  \Sump \rho_p~C_{vp}
     \IntOmegap \chi_p(\Bx)~N_g(\Bx)~N_h(\Bx)~d\Omega ~,
\]
the thermal stiffness matrix as
\[
   K_{gh} := \Sump \kappa_p
      \IntOmegap \chi_p(\Bx)~\Grad{N_g}\cdot\Grad{N_h}~d\Omega ~,
\]
the thermal source vector as
\[
   S_{g} :=  \Sump \LBs \rho_p~\dot{h}_p + \zeta~p_p~\Tr{\Bd^p_p}
        + \zeta~\Bs_p:\Beta^p_p \RBs N_{pg} ~,
\]
and the external heat flux vector as
\[
   q^{\Text}_g := \IntGammaq N_g(\Bx)~\Barq(\Bx)~d\Gamma ~.
\]
Then we get the following semidiscrete system of equations:
\[
    \Sumgh M_{gh}~\Tdot_h + \Sumgh K_{gh}~T_h = S_g + q^{\Text}_g
       ~;~~ g = 1 \dots n_g ~.
\]

\section{Conclusion}
  The derivation described in this paper clearly shows that the integration over the volume
  can be a source of errors in MPM.  Improved algorithms are required.  The stress update algorithm
  can also be improved by using a Lie derivative formulation, possibly with the incorporation of
  a multiplicative decomposition of the deformation gradient.  These issues will be addressed
  in future work.

\section*{Acknowledgments}
  This work was supported by the the U.S. Department of Energy through the
  Center for the Simulation of Accidental Fires and Explosions, under grant
  W-7405-ENG-48.

\bibliographystyle{unsrt}
\bibliography{mybiblio}

\appendix
\section{Useful results}
  {\scriptsize
  \begin{enumerate}
  \item
    {%\Blue
    The integral
    \[
       F(t) = \int_{a(t)}^{b(t)} f(x, t)~\text{dx}
    \]
    is a function of the parameter $t$.  Show that the derivative of $F$
    is given by
    \[
       \Deriv{F}{t} = \Deriv{}{t}\left( \int_{a(t)}^{b(t)} f(x, t)~\text{dx}
           \right) =  \int_{a(t)}^{b(t)} \Partial{f(x, t)}{t}~\text{dx} +
             f[b(t),t]~\Partial{b(t)}{t} - f[a(t),t]~\Partial{a(t)}{t}~.
    \]
    This relation is also known as the {\bf Leibnitz rule}.
    }

    The following proof is taken from ~\cite{Greenberg78}.

    We have,
    \[
        \Deriv{F}{t} = \lim_{\Delt\rightarrow 0}
           \cfrac{F(t + \Delt) - F(t)}{\Delt} ~.
    \] 
    Now,
    \[
      \begin{aligned}
       \cfrac{F(t + \Delt) - F(t)}{\Delt}  & = 
         \cfrac{1}{\Delt} \left[
           \int_{a(t+\Delt)}^{b(t+\Delt)} f(x, t+\Delt)~\text{dx} - 
           \int_{a(t)}^{b(t)} f(x, t)~\text{dx}\right] \\
         & \equiv 
         \cfrac{1}{\Delt} \left[
           \int_{a+\Dela}^{b+\Delb} f(x, t+\Delt)~\text{dx} - 
           \int_{a}^{b} f(x, t)~\text{dx}\right] \\
         & = 
         \cfrac{1}{\Delt} \left[
           -\int_{a}^{a+\Dela} f(x, t+\Delt)~\text{dx} + 
           \int_{a}^{b+\Delb} f(x, t+\Delt)~\text{dx} - 
           \int_{a}^{b} f(x, t)~\text{dx}\right] \\
         & = 
         \cfrac{1}{\Delt} \left[
           -\int_{a}^{a+\Dela} f(x, t+\Delt)~\text{dx} + 
           \int_{a}^{b} f(x, t+\Delt)~\text{dx} + 
           \int_{b}^{b+\Delb} f(x, t+\Delt)~\text{dx} - 
           \int_{a}^{b} f(x, t)~\text{dx}\right] \\
         & = 
           \int_{a}^{b} \cfrac{f(x, t+\Delt) - f(x,t)}{\Delt}~\text{dx} + 
           \cfrac{1}{\Delt}\int_{b}^{b+\Delb} f(x, t+\Delt)~\text{dx} - 
           \cfrac{1}{\Delt}\int_{a}^{a+\Dela} f(x, t+\Delt)~\text{dx} ~.
      \end{aligned}
    \]
    Since $f(x,t)$ is essentially constant over the infinitesimal
    intervals $a < x < a+\Dela$ and $b < x < b+\Delb$, we may write
    \[
       \cfrac{F(t + \Delt) - F(t)}{\Delt}  \approx
           \int_{a}^{b} \cfrac{f(x, t+\Delt) - f(x,t)}{\Delt}~\text{dx} + 
           f(b, t+\Delt)~\cfrac{\Delb}{\Delt} - 
           f(a, t+\Delt)~\cfrac{\Dela}{\Delt}~.
    \]
    Taking the limit as $\Delt\rightarrow 0$, we get
    \[
       \LimDelt \left[\cfrac{F(t + \Delt) - F(t)}{\Delt}\right]  = 
           \LimDelt\left[
           \int_{a}^{b} \cfrac{f(x, t+\Delt) - f(x,t)}{\Delt}~\text{dx}\right] 
           + \LimDelt\left[f(b, t+\Delt)~\cfrac{\Delb}{\Delt}\right] - 
             \LimDelt\left[f(a, t+\Delt)~\cfrac{\Dela}{\Delt}\right]
    \]
    or, 
    \[
       \Deriv{F(t)}{t} = 
             \int_{a(t)}^{b(t)} \Partial{f(x, t)}{t}~\text{dx} +
             f[b(t),t]~\Partial{b(t)}{t} - f[a(t),t]~\Partial{a(t)}{t}~.
       \qquad\qquad\qquad\square
    \]
    
  \item
    {%\Blue
    Let $\Omega(t)$ be a region in Euclidean space with boundary 
    $\Domega(t)$.  Let $\Bx(t)$ be the positions of points in the region
    and let $\Bv(\Bx,t)$ be the velocity field in the region.  
    Let $\Bn(\Bx,t)$ be the outward unit normal to the boundary.  
    Let $\Bf(\Bx,t)$ be a vector field in the region 
    (it may also be a scalar field).  Show that
    \[
      \Deriv{}{t}\left(\IntOmegat \Bf~\DV\right) = 
         \IntOmegat \Partial{\Bf}{t}~\DV + \IntDomegat (\Bv\cdot\Bn)\Bf~\DA ~.
    \]
    This relation is also known as the {\bf Reynold's Transport Theorem} and
    is a generalization of the Leibnitz rule.
    }

    This proof is taken from ~\cite{Belyt00} (also see \cite{Gurtin81}).

    Let $\Omega_0$ be reference configuration of the region $\Omega(t)$.  Let
    the motion and the deformation gradient be given by
    \[
       \Bx = \Bvarphi(\BX, t)~; \qquad\text\qquad 
       \BF(\BX,t) = \GradX{\Bvarphi} ~.
    \]
    Let $J(\BX,t) = \det[\BF(\Bx,t)]$.
    Then, integrals in the current and the reference configurations are
    related by
    \[
       \IntOmegat \Bf(\Bx,t)~\DV = 
          \IntOmegar \Bf[\Bvarphi(\BX,t),t]~J(\BX,t)~\DV =
          \IntOmegar \hat{\Bf}(\BX,t)~J(\BX,t)~\DV ~.
    \]
    The time derivative of an integral over a volume is defined as
    \[
      \Deriv{}{t}\left( \IntOmegat \Bf(\Bx,t)~\DV\right) = 
        \LimDelt \cfrac{1}{\Delt}
         \left(\IntOmegatDelt \Bf(\Bx,t+\Delt)~\DV - 
               \IntOmegat \Bf(\Bx,t)~\DV\right) ~.
    \]
    Converting into integrals over the reference configuration, we get
    \[
      \Deriv{}{t}\left( \IntOmegat \Bf(\Bx,t)~\DV\right) = 
        \LimDelt \cfrac{1}{\Delt}
         \left(\IntOmegar \hat{\Bf}(\BX,t+\Delt)~J(\BX,t+\Delt)~\DV - 
               \IntOmegar \hat{\Bf}(\BX,t)~J(\BX,t)~\DV\right) ~.
    \]
    Since $\Omega_0$ is independent of time, we have
    \[
      \begin{aligned}
      \Deriv{}{t}\left( \IntOmegat \Bf(\Bx,t)~\DV\right) & = 
        \IntOmegar \left[\LimDelt \cfrac{ 
               \hat{\Bf}(\BX,t+\Delt)~J(\BX,t+\Delt) - 
               \hat{\Bf}(\BX,t)~J(\BX,t)}{\Delt} \right]~\DV \\
        & = \IntOmegar \Partial{}{t}[\hat{\Bf}(\BX,t)~J(\BX,t)]~\DV \\
        & = \IntOmegar \left(
              \Partial{}{t}[\hat{\Bf}(\BX,t)]~J(\BX,t)+
              \hat{\Bf}(\BX,t)~\Partial{}{t}[J(\BX,t)]\right) ~\DV 
      \end{aligned}
    \]
    Now, the time derivative of $\det\BF$ is given by 
    (see \cite{Gurtin81}, p. 77)
    \[
       \Partial{J(\BX,t)}{t} = \Partial{}{t}(\det\BF) = (\det\BF)(\Div{\Bv}) 
          = J(\BX,t)~\Div{\Bv}(\Bvarphi(\BX,t),t) 
          = J(\BX,t)~\Div{\Bv}(\Bx,t) ~.
    \]
    Therefore, 
    \[
      \begin{aligned}
      \Deriv{}{t}\left( \IntOmegat \Bf(\Bx,t)~\DV\right) & = 
         \IntOmegar \left(
              \Partial{}{t}[\hat{\Bf}(\BX,t)]~J(\BX,t)+
              \hat{\Bf}(\BX,t)~J(\BX,t)~\Div{\Bv(\Bx,t)}\right) ~\DV \\
         & = 
         \IntOmegar 
              \left(\Partial{}{t}[\hat{\Bf}(\BX,t)]+
              \hat{\Bf}(\BX,t)~\Div{\Bv(\Bx,t)}\right)~J(\BX,t) ~\DV  \\
         & = 
         \IntOmegat 
              \left(\dot{\Bf}(\Bx,t)+
              \Bf(\Bx,t)~\Div{\Bv(\Bx,t)}\right)~\DV 
      \end{aligned}
    \]
    where $\dot{\Bf}$ is the material time derivative of $\Bf$.  Now,
    the material derivative is given by
    \[
      \dot{\Bf}(\Bx,t) = 
        \Partial{\Bf(\Bx,t)}{t} + [\Grad{\Bf(\Bx,t)}]\cdot\Bv(\Bx,t) ~.
    \]
    Therefore,
    \[
      \Deriv{}{t}\left( \IntOmegat \Bf(\Bx,t)~\DV\right) = 
         \IntOmegat 
           \left(
             \Partial{\Bf(\Bx,t)}{t} + [\Grad{\Bf(\Bx,t)}]\cdot\Bv(\Bx,t) +
             \Bf(\Bx,t)~\Div{\Bv(\Bx,t)}\right)~\DV 
    \]
    or,
    \[
      \Deriv{}{t}\left( \IntOmegat \Bf~\DV\right) = 
         \IntOmegat 
           \left(
             \Partial{\Bf}{t} + \Grad{\Bf}\cdot\Bv +
             \Bf~\Div{\Bv}\right)~\DV ~.
    \]
    Using the identity
    \[
       \Div{(\Dyad{\Bv}{\Bw})} = \Bv(\Div{\Bw}) + \Gradv\cdot\Bw 
    \]
    we then have
    \[
      \Deriv{}{t}\left( \IntOmegat \Bf~\DV\right) = 
         \IntOmegat 
           \left(\Partial{\Bf}{t} + \Div{(\Dyad{\Bf}{\Bv})}\right)~\DV ~.
    \]
    Using the divergence theorem and the identity 
    $(\Dyad{\Ba}{\Bb})\cdot\Bn = (\Bb\cdot\Bn)\Ba$ we have
    \[
      \Deriv{}{t}\left( \IntOmegat \Bf~\DV\right) = 
         \IntOmegat\Partial{\Bf}{t}~\DV + 
         \IntDomegat(\Dyad{\Bf}{\Bv})\cdot\Bn~\DV
         = \IntOmegat\Partial{\Bf}{t}~\DV + 
         \IntDomegat(\Bv\cdot\Bn)\Bf~\DV ~.
    \]

  \item
    {%\Blue 
     Show that the balance of mass can be expressed as:
      \[
        \dot{\rho} + \rho~\Div{\Bv} = 0 
      \]
      where $\rho(\Bx,t)$ is the current mass density, $\dot{\rho}$ is
      the material time derivative of $\rho$, and $\Bv(\Bx,t)$ is the 
      velocity of physical particles in the body $\Omega$ bounded by 
      the surface $\Domega$.
    }

    Recall that the general equation for the balance of a physical quantity 
    $f(\Bx,t)$ is given by
    \[
      \Deriv{}{t}\left[\int_{\Omega} f(\Bx,t)~\DV\right] = 
        \int_{\Domega} f(\Bx,t)[u_n(\Bx,t) - \Bv(\Bx,t)\cdot\Bn(\Bx,t)]~\DA + 
        \int_{\Domega} g(\Bx,t)~\DA + \int_{\Omega} h(\Bx,t)~\DV ~.
    \]
    To derive the equation for the balance of mass, we assume that the 
    physical quantity of interest is the mass density $\rho(\Bx,t)$. 
    Since mass is neither created or destroyed, the surface and interior 
    sources are zero, i.e., $g(\Bx, t) = h(\Bx,t) = 0$.  Therefore, we have
    \[
      \Deriv{}{t}\left[\int_{\Omega} \rho(\Bx,t)~\DV\right] = 
        \int_{\Domega} \rho(\Bx,t)[u_n(\Bx,t) - \Bv(\Bx,t)\cdot\Bn(\Bx,t)]~\DA~.
    \]
    Let us assume that the volume $\Omega$ is a control volume (i.e., it 
    does not change with time).  Then the surface $\Domega$ has a zero 
    velocity ($u_n = 0$) and we get
    \[
      \int_{\Omega} \Partial{\rho}{t}~\DV = -
        \int_{\Domega} \rho~(\Bv\cdot\Bn)~\DA~.
    \]
    Using the divergence theorem
    \[
      \int_{\Omega} \Div{\Bv}~\DV = \int_{\Domega} \Bv\cdot\Bn~\DA
    \]
    we get
    \[
      \int_{\Omega} \Partial{\rho}{t}~\DV = -
        \int_{\Omega} \Div{(\rho~\Bv)}~\DV.
    \]
    or,
    \[
      \int_{\Omega} \left[\Partial{\rho}{t} + \Div{(\rho~\Bv)\right]}~\DV = 0 ~.
    \]
    Since $\Omega$ is arbitrary, we must have
    \[
      \Partial{\rho}{t} + \Div{(\rho~\Bv)} = 0 ~.
    \]
    Using the identity
    \[
       \Div{(\varphi~\Bv)} = \varphi~\Div{\Bv} + \Grad{\varphi}\cdot\Bv
    \]
    we have
    \[
      \Partial{\rho}{t} + \rho~\Div{\Bv} + \Grad{\rho}\cdot\Bv = 0 ~.
    \]
    Now, the material time derivative of $\rho$ is defined as
    \[
      \dot{\rho} = \Partial{\rho}{t} + \Grad{\rho}\cdot\Bv ~.
    \]
    Therefore,
    \[
      \dot{\rho} + \rho~\Div{\Bv}  = 0 ~.
    \]
  
  \item
    {%\Blue 
     Show that the balance of linear momentum can be expressed as:
      \[
        \rho~\dot{\Bv} - \Div{\Bsig} - \rho~\Bb = 0 
      \]
      where $\rho(\Bx,t)$ is the mass density, $\Bv(\Bx,t)$ is the velocity,
      $\Bsig(\Bx,t)$ is the Cauchy stress, and $\rho~\Bb$ is the body force 
      density.
    }

    Recall the general equation for the balance of a physical quantity
    \[
      \Deriv{}{t}\left[\int_{\Omega} f(\Bx,t)~\DV\right] = 
        \int_{\Domega} f(\Bx,t)[u_n(\Bx,t) - \Bv(\Bx,t)\cdot\Bn(\Bx,t)]~\DA + 
        \int_{\Domega} g(\Bx,t)~\DA + \int_{\Omega} h(\Bx,t)~\DV ~.
    \]
    In this case the physical quantity of interest is the momentum density,
    i.e., $f(\Bx,t) = \rho(\Bx,t)~\Bv(\Bx,t)$.  The source of momentum flux
    at the surface is the surface traction, i.e., $g(\Bx,t) = \Bt$.  The
    source of momentum inside the body is the body force, i.e., 
    $h(\Bx,t) = \rho(\Bx,t)~\Bb(\Bx,t)$.  Therefore, we have
    \[
      \Deriv{}{t}\left[\int_{\Omega} \rho~\Bv~\DV\right] = 
        \int_{\Domega} \rho~\Bv[u_n - \Bv\cdot\Bn]~\DA + 
        \int_{\Domega} \Bt~\DA + \int_{\Omega} \rho~\Bb~\DV ~.
    \]
    The surface tractions are related to the Cauchy stress by
    \[
       \Bt = \Bsig\cdot\Bn ~.
    \]
    Therefore, 
    \[
      \Deriv{}{t}\left[\int_{\Omega} \rho~\Bv~\DV\right] = 
        \int_{\Domega} \rho~\Bv[u_n - \Bv\cdot\Bn]~\DA + 
        \int_{\Domega} \Bsig\cdot\Bn~\DA + \int_{\Omega} \rho~\Bb~\DV ~.
    \]
    Let us assume that $\Omega$ is an arbitrary fixed control volume.  Then,
    \[
      \int_{\Omega} \Partial{}{t}(\rho~\Bv)~\DV = 
        - \int_{\Domega} \rho~\Bv~(\Bv\cdot\Bn)~\DA + 
        \int_{\Domega} \Bsig\cdot\Bn~\DA + \int_{\Omega} \rho~\Bb~\DV ~.
    \]
    Now, from the definition of the tensor product we have (for all vectors
    $\Ba$)
    \[
       (\Dyad{\Bu}{\Bv})\cdot\Ba = (\Ba\cdot\Bv)~\Bu   ~.
    \]
    Therefore, 
    \[
      \int_{\Omega} \Partial{}{t}(\rho~\Bv)~\DV = 
        - \int_{\Domega} \rho~(\Dyad{\Bv}{\Bv})\cdot\Bn~\DA + 
        \int_{\Domega} \Bsig\cdot\Bn~\DA + \int_{\Omega} \rho~\Bb~\DV ~.
    \]
    Using the divergence theorem
    \[
      \int_{\Omega} \Div{\Bv}~\DV = \int_{\Domega} \Bv\cdot\Bn~\DA
    \]
    we have
    \[
      \int_{\Omega} \Partial{}{t}(\rho~\Bv)~\DV = 
        - \int_{\Omega} \Div{[\rho~(\Dyad{\Bv}{\Bv})]}~\DV + 
        \int_{\Omega} \Div{\Bsig}~\DV + \int_{\Omega} \rho~\Bb~\DV 
    \]
    or, 
    \[
      \int_{\Omega}\left[
        \Partial{}{t}(\rho~\Bv) + \Div{[\Dyad{(\rho~\Bv)}{\Bv}]} - 
        \Div{\Bsig} - \rho~\Bb\right]~\DV = 0 ~.
    \]
    Since $\Omega$ is arbitrary, we have
    \[
      \Partial{}{t}(\rho~\Bv) + \Div{[\Dyad{(\rho~\Bv)}{\Bv}]} - 
      \Div{\Bsig} - \rho~\Bb = 0~.
    \]
    Using the identity
    \[
       \Div{(\Dyad{\Bu}{\Bv})} = (\Div{\Bv})\Bu + (\Gradu)\cdot\Bv
    \]
    we get
    \[
      \Partial{\rho}{t}~\Bv + \rho~\Partial{\Bv}{t} + 
         (\Div{\Bv})(\rho\Bv) + \Grad{(\rho~\Bv)}\cdot\Bv - 
      \Div{\Bsig} - \rho~\Bb = 0 
    \]
    or,
    \[
      \left[\Partial{\rho}{t} + \rho~\Div{\Bv}\right]\Bv + 
        \rho~\Partial{\Bv}{t} + \Grad{(\rho~\Bv)}\cdot\Bv - 
        \Div{\Bsig} - \rho~\Bb = 0 
    \]
    Using the identity 
    \[
      \Grad{(\varphi~\Bv)} = \varphi~\Gradv + \Dyad{\Bv}{(\Grad{\varphi})}
    \]
    we get
    \[
      \left[\Partial{\rho}{t} + \rho~\Div{\Bv}\right]\Bv + 
        \rho~\Partial{\Bv}{t} + 
        \left[\rho~\Gradv + \Dyad{\Bv}{(\Grad{\rho})}\right]\cdot\Bv - 
        \Div{\Bsig} - \rho~\Bb = 0 
    \]
    From the definition
    \[
       (\Dyad{\Bu}{\Bv})\cdot\Ba = (\Ba\cdot\Bv)~\Bu   
    \]
    we have
    \[
       [\Dyad{\Bv}{(\Grad{\rho})}]\cdot\Bv = [\Bv\cdot(\Grad{\rho})]~\Bv   ~.
    \]
    Hence,
    \[
      \left[\Partial{\rho}{t} + \rho~\Div{\Bv}\right]\Bv + 
        \rho~\Partial{\Bv}{t} + 
        \rho~\Gradv\cdot\Bv +  [\Bv\cdot(\Grad{\rho})]~\Bv -
        \Div{\Bsig} - \rho~\Bb = 0 
    \]
    or, 
    \[
      \left[\Partial{\rho}{t} + \Grad{\rho}\cdot\Bv + \rho~\Div{\Bv}\right]\Bv +
        \rho~\Partial{\Bv}{t} + \rho~\Gradv\cdot\Bv -
        \Div{\Bsig} - \rho~\Bb = 0 ~.
    \]
    The material time derivative of $\rho$ is defined as
    \[
      \dot{\rho} = \Partial{\rho}{t} + \Grad{\rho}\cdot\Bv ~.
    \]
    Therefore, 
    \[
      \left[\dot{\rho} + \rho~\Div{\Bv}\right]\Bv +
        \rho~\Partial{\Bv}{t} + \rho~\Gradv\cdot\Bv -
        \Div{\Bsig} - \rho~\Bb = 0 ~.
    \]
    From the balance of mass, we have
    \[
      \dot{\rho} + \rho~\Div{\Bv}  = 0 ~.
    \]
    Therefore, 
    \[
        \rho~\Partial{\Bv}{t} + \rho~\Gradv\cdot\Bv -
        \Div{\Bsig} - \rho~\Bb = 0 ~.
    \]
    The material time derivative of $\Bv$ is defined as
    \[
      \dot{\Bv} = \Partial{\Bv}{t} + \Grad{\Bv}\cdot\Bv ~.
    \]
    Hence,
    \[
        \rho~\dot{\Bv} - \Div{\Bsig} - \rho~\Bb = 0 ~.
    \]
  \item
    {%\Blue 
     Show that the balance of angular momentum can be expressed as:
      \[
        \Bsig  = \Bsig^T
      \]
    }
    We assume that there are no surface couples on $\Domega$ or body couples 
    in $\Omega$.  Recall the general balance equation
    \[
      \Deriv{}{t}\left[\int_{\Omega} f(\Bx,t)~\DV\right] = 
        \int_{\Domega} f(\Bx,t)[u_n(\Bx,t) - \Bv(\Bx,t)\cdot\Bn(\Bx,t)]~\DA + 
        \int_{\Domega} g(\Bx,t)~\DA + \int_{\Omega} h(\Bx,t)~\DV ~.
    \]
    In this case, the physical quantity to be conserved the angular momentum
    density, i.e., $f = \Bcross{\Bx}{(\rho~\Bv)}$. 
    The angular momentum source at the surface is then 
    $g = \Bcross{\Bx}{\Bt}$ and the angular momentum source inside the body 
    is $h = \Bcross{\Bx}{(\rho~\Bb)}$.  The angular momentum and moments are
    calculated with respect to a fixed origin.  Hence we have
    \[
      \Deriv{}{t}\left[\int_{\Omega} \Bcross{\Bx}{(\rho~\Bv)}~\DV\right] = 
        \int_{\Domega} [\Bcross{\Bx}{(\rho~\Bv)}]
           [u_n - \Bv\cdot\Bn]~\DA + 
        \int_{\Domega} \Bcross{\Bx}{\Bt}~\DA + 
        \int_{\Omega} \Bcross{\Bx}{(\rho~\Bb)}~\DV ~.
    \]
    Assuming that $\Omega$ is a control volume, we have
    \[
      \int_{\Omega} \Bcross{\Bx}{\left[\Partial{}{t}(\rho~\Bv)\right]}~\DV = 
        - \int_{\Domega} [\Bcross{\Bx}{(\rho~\Bv)}][\Bv\cdot\Bn]~\DA + 
        \int_{\Domega} \Bcross{\Bx}{\Bt}~\DA + 
        \int_{\Omega} \Bcross{\Bx}{(\rho~\Bb)}~\DV ~.
    \]
    Using the definition of a tensor product we can write
    \[
      [\Bcross{\Bx}{(\rho~\Bv)}][\Bv\cdot\Bn] = 
         [\Dyad{[\Bcross{\Bx}{(\rho~\Bv)}]}{\Bv}]\cdot\Bn ~.
    \]
    Also, $\Bt = \Bsig\cdot\Bn$.  Therefore we have
    \[
      \int_{\Omega} \Bcross{\Bx}{\left[\Partial{}{t}(\rho~\Bv)\right]}~\DV = 
        - \int_{\Domega} [\Dyad{[\Bcross{\Bx}{(\rho~\Bv)}]}{\Bv}]\cdot\Bn ~\DA
        + \int_{\Domega} \Bcross{\Bx}{(\Bsig\cdot\Bn)}~\DA 
        + \int_{\Omega} \Bcross{\Bx}{(\rho~\Bb)}~\DV ~.
    \]
    Using the divergence theorem, we get
    \[
      \int_{\Omega} \Bcross{\Bx}{\left[\Partial{}{t}(\rho~\Bv)\right]}~\DV = 
        - \int_{\Omega} \Div{[\Dyad{[\Bcross{\Bx}{(\rho~\Bv)}]}{\Bv}]}~\DV
        + \int_{\Domega} \Bcross{\Bx}{(\Bsig\cdot\Bn)}~\DA 
        + \int_{\Omega} \Bcross{\Bx}{(\rho~\Bb)}~\DV ~.
    \]
    To convert the surface integral in the above equation into a volume
    integral, it is convenient to use index notation.  Thus,
    \[
       \left[\int_{\Domega} \Bcross{\Bx}{(\Bsig\cdot\Bn)}~\DA\right]_i = 
         \int_{\Domega} e_{ijk}~x_j~\sigma_{kl}~n_l~\DA  = 
         \int_{\Domega} A_{il}~n_l~\DA  = 
         \int_{\Domega} \BA\cdot\Bn~\DA 
    \]
    where $[~]_i$ represents the $i$-th component of the vector.  Using
    the divergence theorem
    \[
       \int_{\Domega} \BA\cdot\Bn~\DA = \int_{\Omega} \Div{\BA}~\DV
         = \int_{\Omega} \Partial{A_{il}}{x_l}~\DV
         = \int_{\Omega} \Partial{}{x_l}(e_{ijk}~x_j~\sigma_{kl})~\DV~.
    \]
    Differentiating, 
    \[
      \begin{aligned}
       \int_{\Domega} \BA\cdot\Bn~\DA 
         & = \int_{\Omega} \left[
             e_{ijk}~\delta_{jl}~\sigma_{kl} +
             e_{ijk}~x_j~\Partial{\sigma_{kl}}{x_l}\right]~\DV
         = \int_{\Omega} \left[
             e_{ijk}~\sigma_{kj} +
             e_{ijk}~x_j~\Partial{\sigma_{kl}}{x_l}\right]~\DV \\
         & = \int_{\Omega} \left[
             e_{ijk}~\sigma_{kj} +
             e_{ijk}~x_j~[\Div{\Bsig}]_l\right]~\DV ~.
      \end{aligned}
    \]
    Expressed in direct tensor notation,
    \[
       \int_{\Domega} \BA\cdot\Bn~\DA 
         = \int_{\Omega} \left[
             [\BCe:\Bsig^T]_i + [\Bcross{\Bx}{(\Div{\Bsig})}]_i\right]~\DV 
    \]
    where $\BCe$ is the third-order permutation tensor.
    Therefore, 
    \[
       \left[\int_{\Domega} \Bcross{\Bx}{(\Bsig\cdot\Bn)}~\DA\right]_i = 
         = \int_{\Omega} \left[
             [\BCe:\Bsig^T]_i + [\Bcross{\Bx}{(\Div{\Bsig})}]_i\right]~\DV 
    \]
    or, 
    \[
       \int_{\Domega} \Bcross{\Bx}{(\Bsig\cdot\Bn)}~\DA = 
         = \int_{\Omega} \left[
             \BCe:\Bsig^T + \Bcross{\Bx}{(\Div{\Bsig})}\right]~\DV ~.
    \]
    The balance of angular momentum can then be written as
    \[
      \int_{\Omega} \Bcross{\Bx}{\left[\Partial{}{t}(\rho~\Bv)\right]}~\DV = 
        - \int_{\Omega} \Div{[\Dyad{[\Bcross{\Bx}{(\rho~\Bv)}]}{\Bv}]}~\DV
        + \int_{\Omega} \left[
             \BCe:\Bsig^T + \Bcross{\Bx}{(\Div{\Bsig})}\right]~\DV 
        + \int_{\Omega} \Bcross{\Bx}{(\rho~\Bb)}~\DV ~.
    \]
    Since $\Omega$ is an arbitrary volume, we have
    \[
      \Bcross{\Bx}{\left[\Partial{}{t}(\rho~\Bv)\right]} = 
        - \Div{[\Dyad{[\Bcross{\Bx}{(\rho~\Bv)}]}{\Bv}]}
        + \BCe:\Bsig^T + \Bcross{\Bx}{(\Div{\Bsig})}
        + \Bcross{\Bx}{(\rho~\Bb)} 
    \]
    or,
    \[
      \Bcross{\Bx}
       {\left[\Partial{}{t}(\rho~\Bv) - \Div{\Bsig} - \rho~\Bb \right]} = 
        - \Div{[\Dyad{[\Bcross{\Bx}{(\rho~\Bv)}]}{\Bv}]}
        + \BCe:\Bsig^T ~.
    \]
    Using the identity, 
    \[
      \Div{(\Dyad{\Bu}{\Bv})} = (\Div{\Bv})\Bu + (\Gradu)\cdot\Bv
    \]
    we get
    \[
      \Div{[\Dyad{[\Bcross{\Bx}{(\rho~\Bv)}]}{\Bv}]} =  
        (\Div{\Bv})[\Bcross{\Bx}{(\rho~\Bv)}] + 
        (\Grad{[\Bcross{\Bx}{(\rho~\Bv)}]})\cdot\Bv ~.
    \]
    The second term on the right can be further simplified using index
    notation as follows.
    \[
      \begin{aligned}
       \left[(\Grad{[\Bcross{\Bx}{(\rho~\Bv)}]})\cdot\Bv\right]_i = 
       \left[(\Grad{[\rho~(\Bcross{\Bx}{\Bv})]})\cdot\Bv\right]_i & = 
         \Partial{}{x_l}(\rho~e_{ijk}~x_j~v_k)~v_l \\
         & = e_{ijk}\left[
               \Partial{\rho}{x_l}~x_j~v_k~v_l  + 
               \rho~\Partial{x_j}{x_l}~v_k~v_l +
               \rho~x_j~\Partial{v_k}{x_l}~v_l\right] \\
         & = (e_{ijk}~x_j~v_k)~\left(\Partial{\rho}{x_l}~v_l\right)  + 
             \rho~(e_{ijk}~\delta_{jl}~v_k~v_l) +
             e_{ijk}~x_j~\left(\rho~\Partial{v_k}{x_l}~v_l\right) \\
         & = [(\Bcross{\Bx}{\Bv})(\Grad{\rho}\cdot\Bv) + 
             \rho~\Bcross{\Bv}{\Bv} + 
             \Bcross{\Bx}{(\rho~\Gradv\cdot\Bv)}]_i \\
         & = [(\Bcross{\Bx}{\Bv})(\Grad{\rho}\cdot\Bv) + 
             \Bcross{\Bx}{(\rho~\Gradv\cdot\Bv)}]_i ~.
      \end{aligned}
    \]
    Therefore we can write
    \[
      \Div{[\Dyad{[\Bcross{\Bx}{(\rho~\Bv)}]}{\Bv}]} =  
        (\rho~\Div{\Bv})(\Bcross{\Bx}{~\Bv}) + 
        (\Grad{\rho}\cdot\Bv)(\Bcross{\Bx}{\Bv}) + 
        \Bcross{\Bx}{(\rho~\Gradv\cdot\Bv)} ~.
    \]
    The balance of angular momentum then takes the form
    \[
      \Bcross{\Bx}
       {\left[\Partial{}{t}(\rho~\Bv) - \Div{\Bsig} - \rho~\Bb \right]} = 
        - (\rho~\Div{\Bv})(\Bcross{\Bx}{~\Bv}) - 
        (\Grad{\rho}\cdot\Bv)(\Bcross{\Bx}{\Bv}) - 
        \Bcross{\Bx}{(\rho~\Gradv\cdot\Bv)} 
        + \BCe:\Bsig^T 
    \]
    or, 
    \[
      \Bcross{\Bx}
       {\left[\Partial{}{t}(\rho~\Bv) + \rho~\Gradv\cdot\Bv 
              - \Div{\Bsig} - \rho~\Bb \right]} = 
        - (\rho~\Div{\Bv})(\Bcross{\Bx}{~\Bv}) - 
        (\Grad{\rho}\cdot\Bv)(\Bcross{\Bx}{\Bv}) 
        + \BCe:\Bsig^T 
    \]
    or,
    \[
      \Bcross{\Bx}
       {\left[\rho\Partial{\Bv}{t} + \Partial{\rho}{t}~\Bv + 
              \rho~\Gradv\cdot\Bv 
              - \Div{\Bsig} - \rho~\Bb \right]} = 
        - (\rho~\Div{\Bv})(\Bcross{\Bx}{~\Bv}) - 
        (\Grad{\rho}\cdot\Bv)(\Bcross{\Bx}{\Bv}) 
        + \BCe:\Bsig^T 
    \]
    The material time derivative of $\Bv$ is defined as
    \[
      \dot{\Bv} = \Partial{\Bv}{t} + \Grad{\Bv}\cdot\Bv ~.
    \]
    Therefore,
    \[
      \Bcross{\Bx}
       {\left[\rho~\dot{\Bv} - \Div{\Bsig} - \rho~\Bb \right]} = 
        - \Bcross{\Bx}{\Partial{\rho}{t}~\Bv} + 
        - (\rho~\Div{\Bv})(\Bcross{\Bx}{~\Bv}) - 
        (\Grad{\rho}\cdot\Bv)(\Bcross{\Bx}{\Bv}) 
        + \BCe:\Bsig^T ~.
    \]
    Also, from the conservation of linear momentum
    \[
        \rho~\dot{\Bv} - \Div{\Bsig} - \rho~\Bb = 0 ~.
    \]
    Hence,
    \[
      \begin{aligned}
        0 & = \Bcross{\Bx}{\Partial{\rho}{t}~\Bv} + 
        (\rho~\Div{\Bv})(\Bcross{\Bx}{~\Bv}) + 
        (\Grad{\rho}\cdot\Bv)(\Bcross{\Bx}{\Bv}) 
        - \BCe:\Bsig^T \\
          & = \left(\Partial{\rho}{t} + \rho\Div{\Bv} + 
                     \Grad{\rho}\cdot\Bv \right)(\Bcross{\Bx}{\Bv}) 
        - \BCe:\Bsig^T ~.
      \end{aligned}
    \]
    The material time derivative of $\rho$ is defined as
    \[
      \dot{\rho} = \Partial{\rho}{t} + \Grad{\rho}\cdot\Bv ~.
    \]
    Hence,
    \[
      (\dot{\rho} + \rho~\Div{\Bv})(\Bcross{\Bx}{\Bv})
        - \BCe:\Bsig^T = 0 ~.
    \]
    From the balance of mass
    \[
      \dot{\rho} + \rho~\Div{\Bv} = 0 ~.
    \]
    Therefore,
    \[
       \BCe:\Bsig^T = 0 ~.
    \]
    In index notation,
    \[
       e_{ijk}~\sigma_{kj} = 0 ~.
    \]
    Expanding out, we get
    \[
       \sigma_{12} - \sigma_{21} = 0 ~;~~
       \sigma_{23} - \sigma_{32} = 0 ~;~~
       \sigma_{31} - \sigma_{13} = 0 ~.
    \]
    Hence, 
    \[
       \Bsig = \Bsig^T
    \]
  
  \item
    {%\Blue 
     Show that the balance of energy can be expressed as:
      \[
        \rho~\dot{e} - \Bsig:(\Gradv) + \Div{\Bq} - \rho~s  = 0
      \]
      where $\rho(\Bx,t)$ is the mass density, $e(\Bx,t)$ is the internal energy
      per unit mass, $\Bsig(\Bx,t)$ is the Cauchy stress, $\Bv(\Bx,t)$ is
      the particle velocity, $\Bq$ is the heat flux vector, and $s$ is the
      rate at which energy is generated by sources inside the volume
      (per unit mass).
    }

    Recall the general balance equation
    \[
      \Deriv{}{t}\left[\int_{\Omega} f(\Bx,t)~\DV\right] = 
        \int_{\Domega} f(\Bx,t)[u_n(\Bx,t) - \Bv(\Bx,t)\cdot\Bn(\Bx,t)]~\DA + 
        \int_{\Domega} g(\Bx,t)~\DA + \int_{\Omega} h(\Bx,t)~\DV ~.
    \]
    In this case, the physical quantity to be conserved the total energy density
    which is the sum of the internal energy density and the kinetic energy 
    density, i.e., $f = \rho~e + 1/2~\rho~|\Bv\cdot\Bv|$.
    The energy source at the surface is a sum of the rate of work done by
    the applied tractions and the rate of heat leaving the volume 
    (per unit area), i.e, $g = \Bv\cdot\Bt - \Bq\cdot\Bn$ where $\Bn$ is the 
    outward unit normal to the surface.  The energy source inside the body 
    is the sum of the rate of work done by the body forces and the rate of
    energy generated by internal sources, i.e., 
    $h = \Bv\cdot(\rho\Bb) + \rho~s$.  

    Hence we have
    \[
      \Deriv{}{t}\left[\int_{\Omega}
         \rho~\left(e + \Half~\Bv\cdot\Bv\right)~\DV\right] = 
      \int_{\Domega}\rho~\left(e + \Half~\Bv\cdot\Bv\right)
            (u_n - \Bv\cdot\Bn)~\DA + 
      \int_{\Domega} (\Bv\cdot\Bt - \Bq\cdot\Bn)~\DA + 
      \int_{\Omega} \rho~(\Bv\cdot\Bb + s)~\DV ~.
    \]
    Let $\Omega$ be a control volume that does not change with time.  Then
    we get
    \[
      \int_{\Omega}
        \Partial{}{t}\left[\rho~\left(e + \Half~\Bv\cdot\Bv\right)\right]~\DV = 
      -\int_{\Domega}\rho~\left(e + \Half~\Bv\cdot\Bv\right)
           (\Bv\cdot\Bn)~\DA + 
      \int_{\Domega} (\Bv\cdot\Bt - \Bq\cdot\Bn)~\DA + 
      \int_{\Omega} \rho~(\Bv\cdot\Bb + s)~\DV ~.
    \]
    Using the relation $\Bt = \Bsig\cdot\Bn$, the identity
    $\Bv\cdot(\Bsig\cdot\Bn) = (\Bsig^T\cdot\Bv)\cdot\Bn$, and invoking
    the symmetry of the stress tensor, we get
    \[
      \int_{\Omega}
        \Partial{}{t}\left[\rho~\left(e + \Half~\Bv\cdot\Bv\right)\right]~\DV = 
      -\int_{\Domega}\rho~\left(e + \Half~\Bv\cdot\Bv\right)
           (\Bv\cdot\Bn)~\DA + 
      \int_{\Domega} (\Bsig\cdot\Bv - \Bq)\cdot\Bn~\DA + 
      \int_{\Omega} \rho~(\Bv\cdot\Bb + s)~\DV ~.
    \]
    We now apply the divergence theorem to the surface integrals to get
    \[
      \int_{\Omega}
        \Partial{}{t}\left[\rho~\left(e + \Half~\Bv\cdot\Bv\right)\right]~\DV = 
      -\int_{\Omega}
        \Div{\left[\rho~\left(e + \Half~\Bv\cdot\Bv\right)\Bv\right]}~\DA 
      +\int_{\Omega} \Div{(\Bsig\cdot\Bv)}~\DA 
      - \int_{\Omega} \Div{\Bq}~\DA 
      +\int_{\Omega} \rho~(\Bv\cdot\Bb + s)~\DV ~.
    \]
    Since $\Omega$ is arbitrary, we have
    \[
      \Partial{}{t}\left[\rho~\left(e + \Half~\Bv\cdot\Bv\right)\right] = 
      - \Div{\left[\rho~\left(e + \Half~\Bv\cdot\Bv\right)\Bv\right]}
      + \Div{(\Bsig\cdot\Bv)}
      - \Div{\Bq}
      + \rho~(\Bv\cdot\Bb + s)~.
    \]
    Expanding out the left hand side, we have
    \[
      \begin{aligned}
        \Partial{}{t}\left[\rho~\left(e + \Half~\Bv\cdot\Bv\right)\right] & = 
          \Partial{\rho}{t}\left(e + \Half~\Bv\cdot\Bv\right) + 
          \rho~\left(\Partial{e}{t} + \Half~\Partial{}{t}(\Bv\cdot\Bv)\right) \\
           & = 
          \Partial{\rho}{t}\left(e + \Half~\Bv\cdot\Bv\right) + 
          \rho~\Partial{e}{t} + \rho~\Partial{\Bv}{t}\cdot\Bv ~.
      \end{aligned}
    \]
    For the first term on the right hand side, we use the identity
    $\Div{(\varphi~\Bv)} = \varphi~\Div{\Bv} + \Grad{\varphi}\cdot\Bv$ to get
    \[
      \begin{aligned}
        \Div{\left[\rho~\left(e + \Half~\Bv\cdot\Bv\right)\Bv\right]} & = 
        \rho~\left(e + \Half~\Bv\cdot\Bv\right)~\Div{\Bv} + 
        \Grad{\left[\rho~\left(e + \Half~\Bv\cdot\Bv\right)\right]}\cdot\Bv \\
        & = \rho~\left(e + \Half~\Bv\cdot\Bv\right)~\Div{\Bv} + 
        \left(e + \Half~\Bv\cdot\Bv\right)\Grad{\rho}\cdot\Bv +
        \rho~\Grad{\left(e + \Half~\Bv\cdot\Bv\right)}\cdot\Bv \\
        & = \rho~\left(e + \Half~\Bv\cdot\Bv\right)~\Div{\Bv} + 
        \left(e + \Half~\Bv\cdot\Bv\right)\Grad{\rho}\cdot\Bv +
        \rho~\Grad{e}\cdot{\Bv} + \Half~\rho~\Grad{(\Bv\cdot\Bv)}\cdot\Bv \\
        & = \rho~\left(e + \Half~\Bv\cdot\Bv\right)~\Div{\Bv} + 
        \left(e + \Half~\Bv\cdot\Bv\right)\Grad{\rho}\cdot\Bv +
        \rho~\Grad{e}\cdot{\Bv} + \rho~(\Gradv^T\cdot\Bv)\cdot\Bv \\
        & = \rho~\left(e + \Half~\Bv\cdot\Bv\right)~\Div{\Bv} + 
        \left(e + \Half~\Bv\cdot\Bv\right)\Grad{\rho}\cdot\Bv +
        \rho~\Grad{e}\cdot{\Bv} + \rho~(\Gradv\cdot\Bv)\cdot\Bv ~.
      \end{aligned}
    \]
    For the second term on the right we use the identity 
    $\Div{(\BS^T\cdot\Bv)} = \BS:\Gradv + (\Div{\BS})\cdot\Bv$ and the
    symmetry of the Cauchy stress tensor to get
    \[
      \Div{(\Bsig\cdot\Bv)} = \Bsig:\Gradv + (\Div{\Bsig})\cdot\Bv ~.
    \]
    After collecting terms and rearranging, we get
    \[
       \begin{aligned}
       & \left(\Partial{\rho}{t}+ \rho~\Div{\Bv} + \Grad{\rho}\cdot\Bv\right)
          \left(e + \Half~\Bv\cdot\Bv\right) + 
       \left(\rho~\Partial{\Bv}{t} + \rho~\Gradv\cdot\Bv - 
            \Div{\Bsig} - \rho~\Bb\right)\cdot\Bv +
       \rho~\left(\Partial{e}{t} + \Grad{e}\cdot{\Bv}\right) + \\
       & \qquad - \Bsig:\Gradv 
       + \Div{\Bq}
       - \rho~s = 0 ~.
       \end{aligned}
    \]
    Applying the balance of mass to the first term and the balance of linear
    momentum to the second term, and using the material time derivative of
    the internal energy
    \[
      \dot{e} = \Partial{e}{t} + \Grad{e}\cdot\Bv 
    \]
    we get the final form of the balance of energy:
    \[
       \rho~\dot{e} - \Bsig:\Gradv + \Div{\Bq} - \rho~s = 0 ~.
    \]

  \item
    {%\Blue 
     Show that the Clausius-Duhem inequality in integral form:
       \[
       \Deriv{}{t}\left(\IntOmega \rho~\eta~\DV\right) \ge
       \IntDomega \rho~\eta~(u_n - \Bv\cdot\Bn)~\DA - 
       \IntDomega \cfrac{\Bq\cdot\Bn}{T}~\DA + 
        \IntOmega \cfrac{\rho~s}{T}~\DV ~.
       \]
       can be written in differential form as
       \[
         \rho~\dot{\eta} \ge - \Div{\left(\cfrac{\Bq}{T}\right)} 
            + \cfrac{\rho~s}{T} ~.
       \]
     }
     Assume that $\Omega$ is an arbitrary fixed control volume.  Then
     $u_n = 0$ and the derivative can be take inside the integral to give
     \[
       \IntOmega \Partial{}{t}(\rho~\eta)~\DV \ge
       -\IntDomega \rho~\eta~(\Bv\cdot\Bn)~\DA - 
       \IntDomega \cfrac{\Bq\cdot\Bn}{T}~\DA + 
        \IntOmega \cfrac{\rho~s}{T}~\DV ~.
     \]
     Using the divergence theorem, we get
     \[
       \IntOmega \Partial{}{t}(\rho~\eta)~\DV \ge
       -\IntOmega \Div{(\rho~\eta~\Bv)}~\DV - 
        \IntOmega \Div{\left(\cfrac{\Bq}{T}\right)}~\DV + 
        \IntOmega \cfrac{\rho~s}{T}~\DV ~.
     \]
     Since $\Omega$ is arbitrary, we must have
     \[
       \Partial{}{t}(\rho~\eta) \ge
       -\Div{(\rho~\eta~\Bv)} - 
        \Div{\left(\cfrac{\Bq}{T}\right)} + 
        \cfrac{\rho~s}{T} ~.
     \]
     Expanding out
     \[
       \Partial{\rho}{t}~\eta + \rho~\Partial{\eta}{t}  \ge
       -\Grad{(\rho_\eta)}\cdot\Bv - \rho~\eta~(\Div{\Bv}) -
        \Div{\left(\cfrac{\Bq}{T}\right)} + 
        \cfrac{\rho~s}{T} 
     \]
     or, 
     \[
       \Partial{\rho}{t}~\eta + \rho~\Partial{\eta}{t}  \ge
       -\eta~\Grad{\rho}\cdot\Bv - \rho~\Grad{\eta}\cdot\Bv - 
        \rho~\eta~(\Div{\Bv}) -
        \Div{\left(\cfrac{\Bq}{T}\right)} + 
        \cfrac{\rho~s}{T} 
     \]
     or, 
     \[
       \left(\Partial{\rho}{t} + \Grad{\rho}\cdot\Bv + \rho~\Div{\Bv}\right)
       ~\eta +
       \rho~\left(\Partial{\eta}{t} + \Grad{\eta}\cdot\Bv\right)
       \ge -\Div{\left(\cfrac{\Bq}{T}\right)} + 
        \cfrac{\rho~s}{T} ~.
     \]
     Now, the material time derivatives of $\rho$ and $\eta$ are given by
     \[
       \dot{\rho} = \Partial{\rho}{t} + \Grad{\rho}\cdot\Bv ~;~~
       \dot{\eta} = \Partial{\eta}{t} + \Grad{\eta}\cdot\Bv ~.
     \]
     Therefore, 
     \[
       \left(\dot{\rho} + \rho~\Div{\Bv}\right)~\eta +
       \rho~\dot{\eta}
       \ge -\Div{\left(\cfrac{\Bq}{T}\right)} + 
        \cfrac{\rho~s}{T} ~.
     \]
     From the conservation of mass $\dot{\rho} + \rho~\Div{\Bv} = 0$.  Hence,
     \[
       \rho~\dot{\eta} \ge -\Div{\left(\cfrac{\Bq}{T}\right)} + 
        \cfrac{\rho~s}{T} ~.
     \]
   
  \item
    {%\Blue 
    Show that the Clausius-Duhem inequality 
       \[
         \rho~\dot{\eta} \ge - \Div{\left(\cfrac{\Bq}{T}\right)} 
            + \cfrac{\rho~s}{T} 
       \]
       can be expressed in terms of the internal energy as
       \[
         \rho~(\dot{e} - T~\dot{\eta}) - \Bsig:\Gradv \le 
           - \cfrac{\Bq\cdot\Grad{T}}{T} ~.
       \]
     }
     Using the identity
     $ \Div{(\varphi~\Bv)} = \varphi~\Div{\Bv} + \Bv\cdot\Grad{\varphi}$
     in the Clausius-Duhem inequality, we get
     \[
        \rho~\dot{\eta}  \ge  - \Div{\left(\cfrac{\Bq}{T}\right)} 
            + \cfrac{\rho~s}{T}  \qquad\Tor\qquad
        \rho~\dot{\eta}  \ge - \cfrac{1}{T}~\Div{\Bq} - 
               \Bq\cdot\Grad{\left(\cfrac{1}{T}\right)}
            + \cfrac{\rho~s}{T} ~. 
     \]
     Now, using index notation with respect to a Cartesian basis $\Be_j$, 
     \[
       \Grad{\left(\cfrac{1}{T}\right)} = 
         \Partial{}{x_j}\left(T^{-1}\right)~\Be_j = 
         -\left(T^{-2}\right)~\Partial{T}{x_j}~\Be_j
         = -\cfrac{1}{T^2}~\Grad{T} ~.
     \]
     Hence,
     \[
       \rho~\dot{\eta} \ge - \cfrac{1}{T}~\Div{\Bq} + 
               \cfrac{1}{T^2}~\Bq\cdot\Grad{T}
            + \cfrac{\rho~s}{T} \qquad\Tor\qquad
       \rho~\dot{\eta} \ge -\cfrac{1}{T}\left(\Div{\Bq} - \rho~s\right) + 
               \cfrac{1}{T^2}~\Bq\cdot\Grad{T} ~.
     \]
     Recall the balance of energy
     \[
       \rho~\dot{e} - \Bsig:\Gradv + \Div{\Bq} - \rho~s = 0 
       \qquad \implies \qquad
       \rho~\dot{e} - \Bsig:\Gradv = - (\Div{\Bq} - \rho~s) ~.
     \]
     Therefore, 
     \[
       \rho~\dot{\eta} \ge \cfrac{1}{T}\left(\rho~\dot{e}-\Bsig:\Gradv\right) + 
               \cfrac{1}{T^2}~\Bq\cdot\Grad{T} 
       \qquad \implies \qquad
       \rho~\dot{\eta}~T \ge \rho~\dot{e}-\Bsig:\Gradv + 
               \cfrac{\Bq\cdot\Grad{T}}{T} ~. 
     \]
     Rearranging, 
     \[
         \rho~(\dot{e} - T~\dot{\eta}) - \Bsig:\Gradv \le 
           - \cfrac{\Bq\cdot\Grad{T}}{T} ~.
     \]

  \item
    {%\Blue 
       For thermoelastic materials, the internal energy is a function 
       only of the deformation gradient and the temperature, i.e., 
       $e = e(\BF, T)$.  Show that, for thermoelastic materials, the 
       Clausius-Duhem inequality
       \[
         \rho~(\dot{e} - T~\dot{\eta}) - \Bsig:\Gradv \le 
           - \cfrac{\Bq\cdot\Grad{T}}{T} 
       \]
       can be expressed as
       \[
         \rho~\left(\Partial{e}{\eta} - T\right)~\dot{\eta} +
              \left(\rho~\Partial{e}{\BF} - \Bsig\cdot\BF^{-T}\right):\BFdot 
          \le - \cfrac{\Bq\cdot\Grad{T}}{T} ~.
       \]
     }
     Since $e = e(\BF, T)$, we have
     \[
        \dot{e} = \Partial{e}{\BF}:\BFdot + \Partial{e}{\eta}~\dot{\eta} ~.
     \]
     Therefore, 
     \[
       \rho~\left(\Partial{e}{\BF}:\BFdot + \Partial{e}{\eta}~\dot{\eta} 
          - T~\dot{\eta}\right) - \Bsig:\Gradv \le 
           - \cfrac{\Bq\cdot\Grad{T}}{T} 
       \qquad\Tor\qquad
       \rho\left(\Partial{e}{\eta} - T\right)~\dot{\eta} + 
       \rho~\Partial{e}{\BF}:\BFdot  
       - \Bsig:\Gradv \le - \cfrac{\Bq\cdot\Grad{T}}{T} ~.
     \]
     Now, $\Gradv = \Bl = \BFdot\cdot\BF^{-1}$.  Therefore, using the 
     identity $\BA:(\BB\cdot\BC) = (\BA\cdot\BC^T):\BB$, we have
     \[
        \Bsig:\Gradv = \Bsig:(\BFdot\cdot\BF^{-1})
                     = (\Bsig\cdot\BF^{-T}):\BFdot ~.
     \]
     Hence,
     \[
       \rho\left(\Partial{e}{\eta} - T\right)~\dot{\eta} + 
       \rho~\Partial{e}{\BF}:\BFdot  
       - (\Bsig\cdot\BF^{-T}):\BFdot \le - \cfrac{\Bq\cdot\Grad{T}}{T} 
     \]
     or,
     \[
         \rho~\left(\Partial{e}{\eta} - T\right)~\dot{\eta} +
              \left(\rho~\Partial{e}{\BF} - \Bsig\cdot\BF^{-T}\right):\BFdot 
          \le - \cfrac{\Bq\cdot\Grad{T}}{T} ~.
     \]

  \item
    {%\Blue 
       Show that, for thermoelastic materials, the balance of energy 
       \[
         \rho~\dot{e} - \Bsig:\Gradv + \Div{\Bq} - \rho~s = 0 ~.
       \]
       can be expressed as
       \[
         \rho~T~\dot{\eta} = - \Div{\Bq} + \rho~s ~.
       \]
     }
     Since $e = e(\BF, T)$, we have
     \[
        \dot{e} = \Partial{e}{\BF}:\BFdot + \Partial{e}{\eta}~\dot{\eta} ~.
     \]
     Plug into energy equation to get
     \[
       \rho~\Partial{e}{\BF}:\BFdot + \rho~\Partial{e}{\eta}~\dot{\eta}
          - \Bsig:\Gradv + \Div{\Bq} - \rho~s = 0 ~.
     \]
     Recall, 
     \[ 
        \Partial{e}{\eta} = T \qquad\text{and}\qquad
        \rho~\Partial{e}{\BF} = \Bsig\cdot\BF^{-T} ~.
     \]
     Hence,
     \[
       (\Bsig\cdot\BF^{-T}):\BFdot + \rho~T~\dot{\eta}
          - \Bsig:\Gradv + \Div{\Bq} - \rho~s = 0 ~.
     \]
     Now, $\Gradv = \Bl = \BFdot\cdot\BF^{-1}$.  Therefore, using the 
     identity $\BA:(\BB\cdot\BC) = (\BA\cdot\BC^T):\BB$, we have
     \[
        \Bsig:\Gradv = \Bsig:(\BFdot\cdot\BF^{-1})
                     = (\Bsig\cdot\BF^{-T}):\BFdot ~.
     \]
     Plugging into the energy equation, we have
     \[
       \Bsig:\Gradv + \rho~T~\dot{\eta}
          - \Bsig:\Gradv + \Div{\Bq} - \rho~s = 0 
     \]
     or,
     \[
         \rho~T~\dot{\eta} = - \Div{\Bq} + \rho~s ~.
     \]

  \item 
    {%\Blue 
      Show that, for thermoelastic materials, the Cauchy stress can 
      be expressed in terms of the Green strain as
      \[
        \Bsig = \rho~\BF\cdot\Partial{e}{\BE}\cdot\BF^T ~.
      \]
    }
    Recall that the Cauchy stress is given by
    \[
      \Bsig = \rho~\Partial{e}{\BF}\cdot\BF^T 
      \qquad \implies \qquad
      \sigma_{ij} = \rho~\Partial{e}{F_{ik}}F^T_{kj} 
                  = \rho~\Partial{e}{F_{ik}}F_{jk}  ~.
    \]
    The Green strain $\BE = \BE(\BF) = \BE(\BU)$ and 
    $e = e(\BF,\eta) = e(\BU,\eta)$.  Hence, using the chain rule,
    \[
      \Partial{e}{\BF} = \Partial{e}{\BE}:\Partial{\BE}{\BF}
      \qquad \implies \qquad
      \Partial{e}{F_{ik}} = \Partial{e}{E_{lm}}~\Partial{E_{lm}}{F_{ik}} ~.
    \]
    Now,
    \[
      \BE = \Half(\BF^T\cdot\BF - \Bone) 
      \qquad \implies \qquad
      E_{lm} = \Half(F^T_{lp}~F_{pm} - \delta_{lm}) 
             = \Half(F_{pl}~F_{pm} - \delta_{lm}) ~.
    \]
    Taking the derivative with respect to $\BF$, we get
    \[
      \Partial{\BE}{\BF} = \Half\left(\Partial{\BF^T}{\BF}\cdot\BF +
           \BF^T\cdot\Partial{\BF}{\BF}\right)
      \qquad \implies \qquad
      \Partial{E_{lm}}{F_{ik}} = \Half\left(\Partial{F_{pl}}{F_{ik}}~F_{pm} +
           F_{pl}~\Partial{F_{pm}}{F_{ik}}\right) ~.
    \]
    Therefore,
    \[
      \Bsig = \Half~\rho~\left[\Partial{e}{\BE}:
          \left(\Partial{\BF^T}{\BF}\cdot\BF +
           \BF^T\cdot\Partial{\BF}{\BF}\right)\right]\cdot\BF^T 
      \qquad \implies \qquad
      \sigma_{ij} = \Half~\rho~\left[\Partial{e}{E_{lm}}
          \left(\Partial{F_{pl}}{F_{ik}}~F_{pm} +
           F_{pl}~\Partial{F_{pm}}{F_{ik}}\right)\right]~F_{jk} ~.
    \]
    Recall,
    \[
      \Partial{\BA}{\BA} \equiv \Partial{A_{ij}}{A_{kl}} = 
      \delta_{ik}~\delta_{jl} \qquad \text{and} \qquad
      \Partial{\BA^T}{\BA} \equiv \Partial{A_{ji}}{A_{kl}} = 
      \delta_{jk}~\delta_{il} ~.
    \]
    Therefore,
    \[
      \sigma_{ij} = \Half~\rho~\left[\Partial{e}{E_{lm}}
          \left(\delta_{pi}~\delta_{lk}~F_{pm} +
           F_{pl}~\delta_{pi}~\delta_{mk}\right)\right]~F_{jk} 
       = \Half~\rho~\left[\Partial{e}{E_{lm}}
          \left(\delta_{lk}~F_{im} +
           F_{il}~\delta_{mk}\right)\right]~F_{jk} 
    \]
    or, 
    \[
      \sigma_{ij} = \Half~\rho~\left[\Partial{e}{E_{km}}~F_{im} +
           \Partial{e}{E_{lk}}~F_{il}\right]~F_{jk} 
      \qquad \implies \qquad
      \Bsig = \Half~\rho~\left[\BF\cdot\left(\Partial{e}{\BE}\right)^T +
           \BF\cdot\Partial{e}{\BE}\right]\cdot\BF^T 
    \]
    or,
    \[
      \Bsig = \Half~\rho~\BF\cdot\left[\left(\Partial{e}{\BE}\right)^T +
           \Partial{e}{\BE}\right]\cdot\BF^T ~.
    \]
    From the symmetry of the Cauchy stress, we have
    \[
       \Bsig = (\BF\cdot\BA)\cdot\BF^T \qquad \text{and} \qquad
       \Bsig^T = \BF\cdot(\BF\cdot\BA)^T = \BF\cdot\BA^T\cdot\BF^T 
       \qquad \text{and} \qquad \Bsig = \Bsig^T \implies \BA = \BA^T ~.
    \]
    Therefore, 
    \[
       \Partial{e}{\BE} = \left(\Partial{e}{\BE}\right)^T
    \]
    and we get
    \[
      \Bsig = ~\rho~\BF\cdot\Partial{e}{\BE}\cdot\BF^T ~.
    \]
  
  \item 
    {%\Blue 
      For thermoelastic materials, the specific internal energy 
      is given by 
      \[
         e = \bar{e}(\BE, \eta)
      \]
      where $\BE$ is the Green strain and $\eta$ is the specific entropy.  
      Show that
      \[
      \Deriv{}{t}(e - T~\eta) = - \dot{T}~\eta + \cfrac{1}{\rho_0}~\BS:\BEdot
      \qquad\text{and}\qquad
      \Deriv{}{t}(e - T~\eta - \cfrac{1}{\rho_0}~\BS:\BE) = 
        - \dot{T}~\eta - \cfrac{1}{\rho_0}~\BSdot:\BE 
      \]
      where $\rho_0$ is the initial density, $T$ is the absolute temperature,
      $\BS$ is the 2nd Piola-Kirchhoff
      stress, and a dot over a quantity indicates the material time derivative.
    }

    Taking the material time derivative of the specific internal energy, we
    get 
    \[
       \dot{e} = \Partial{\bar{e}}{\BE}:\BEdot + 
                 \Partial{\bar{e}}{\eta}~\Etadot ~.
    \]
    Now, for thermoelastic materials,
    \[
       T = \Partial{\bar{e}}{\eta} \qquad \text{and} \qquad 
       \BS = \rho_0~\Partial{\bar{e}}{\BE} ~.
    \]
    Therefore, 
    \[
       \dot{e} = \cfrac{1}{\rho_0}~\BS:\BEdot + T~\Etadot ~.
       \qquad \implies \qquad
       \dot{e} - T~\Etadot = \cfrac{1}{\rho_0}~\BS:\BEdot ~.
    \]
    Now,
    \[
       \Deriv{}{t}(T~\eta) = \Tdot~\eta + T~\Etadot ~.
    \]
    Therefore, 
    \[
       \dot{e} - \Deriv{}{t}(T~\eta) + \Tdot~\eta 
          = \cfrac{1}{\rho_0}~\BS:\BEdot 
       \qquad \implies \qquad
       \Deriv{}{t}(e - T~\eta) = -\Tdot~\eta + 
           \cfrac{1}{\rho_0}~\BS:\BEdot ~.
    \]
    Also,
    \[
       \Deriv{}{t}\left(\cfrac{1}{\rho_0}~\BS:\BE\right) = 
         \cfrac{1}{\rho_0}~\BS:\BEdot + \cfrac{1}{\rho_0}~\BSdot:\BE ~.
    \]
    Hence, 
    \[
       \dot{e} - \Deriv{}{t}(T~\eta) + \Tdot~\eta 
          = \Deriv{}{t}\left(\cfrac{1}{\rho_0} \BS:\BE\right) - 
            \cfrac{1}{\rho_0}~\BSdot:\BE 
       \qquad \implies \qquad
       \Deriv{}{t}\left(e - T~\eta - \cfrac{1}{\rho_0}~\BS:\BE\right)
         = - \Tdot~\eta - \cfrac{1}{\rho_0}~\BSdot:\BE ~.
    \]

  \item 
    {%\Blue 
      For thermoelastic materials, show that the following relations
      hold:
      \[
        \Partial{\psi}{\BE}  = \cfrac{1}{\rho_0}~\hat{\BS}(\BE,T) ~;~~
        \Partial{\psi}{T}  = -\hat{\eta}(\BE,T) ~;~~
        \Partial{g}{\BS}  = \cfrac{1}{\rho_0}~\tilde{\BE}(\BS, T) ~;~~
        \Partial{g}{T}  = \tilde{\eta}(\BS, T) 
      \]
      where $\psi(\BE,T)$ is the Helmholtz free energy and $g(\BS,T)$ is the 
      Gibbs free energy.

      Also show that
      \[
        \Partial{\hat{\BS}}{T} = - \rho_0~\Partial{\hat{\eta}}{\BE} 
        \qquad\text{and}\qquad
        \Partial{\tilde{\BE}}{T} = \rho_0~\Partial{\tilde{\eta}}{\BS} ~.
      \]
    }

    Recall that
    \[
      \psi(\BE, T) = e - T~\eta = \bar{e}(\BE, \eta) - T~\eta ~.
    \]
    and
    \[
      g(\BS, T) = - e + T~\eta + \cfrac{1}{\rho_0}~\BS:\BE ~.
    \]
    (Note that we can choose any functional dependence that we like, because
    the quantities $e$, $\eta$, $\BE$ are the actual quantities and not any
    particular functional relations).

    The derivatives are
    \[
      \Partial{\psi}{\BE} = \Partial{\bar{e}}{\BE} = \cfrac{1}{\rho_0}~\BS
      ~;\qquad
      \Partial{\psi}{T} = - \eta
      ~.
    \]
    and
    \[
      \Partial{g}{\BS} = \cfrac{1}{\rho_0}~\Partial{\BS}{\BS}:\BE 
        = \cfrac{1}{\rho_0}~\BE
      ~;\qquad
      \Partial{g}{T} =  \eta ~.
    \]
    Hence,
    \[
      \Partial{\psi}{\BE}  = \cfrac{1}{\rho_0}~\hat{\BS}(\BE,T) ~;~~
      \Partial{\psi}{T}  = -\hat{\eta}(\BE,T) ~;~~
      \Partial{g}{\BS}  = \cfrac{1}{\rho_0}~\tilde{\BE}(\BS, T) ~;~~
      \Partial{g}{T}  = \tilde{\eta}(\BS, T) 
    \]

    From the above, we have
    \[
       \PPartialA{\psi}{T}{\BE} = \PPartialA{\psi}{\BE}{T} 
       \qquad\implies\qquad
       -\Partial{\hat{\eta}}{\BE} = \cfrac{1}{\rho_0}\Partial{\hat{\BS}}{T} ~.
    \]
    and
    \[
       \PPartialA{g}{T}{\BS} = \PPartialA{g}{\BS}{T} 
       \qquad\implies\qquad
       \Partial{\tilde{\eta}}{\BS} = 
        \cfrac{1}{\rho_0}\Partial{\tilde{\BE}}{T} ~.
    \]
    Hence,
    \[
        \Partial{\hat{\BS}}{T} = - \rho_0~\Partial{\hat{\eta}}{\BE} 
        \qquad\text{and}\qquad
        \Partial{\tilde{\BE}}{T} = \rho_0~\Partial{\tilde{\eta}}{\BS} ~.
    \]

    \item
    {%\Blue 
      For thermoelastic materials, show that the following relations
      hold:
      \[
         \Partial{\hat{e}(\BE,T)}{T} =  T~\Partial{\hat{\eta}}{T} = 
           -T~\PPartial{\hat{\psi}}{T} 
      \]
      and
      \[
         \Partial{\tilde{e}(\BS,T)}{T} =  
           T~\Partial{\tilde{\eta}}{T} + 
           \cfrac{1}{\rho_0}~\BS:\Partial{\tilde{\BE}}{T} = 
           T~\PPartial{\tilde{g}}{T}  
           + \BS:\PPartialA{\tilde{g}}{\BS}{T}~.
      \]
    }
    Recall, 
    \[
       \hat{\psi}(\BE,T) = 
       \psi = e - T~\eta = \hat{e}(\BE, T) - T~\hat{\eta}(\BE, T) 
    \]
    and
    \[
       \tilde{g}(\BS,T) = 
       g = - e + T~\eta + \cfrac{1}{\rho_0}~\BS:\BE 
                 = -\tilde{e}(\BS, T) + T~\tilde{\eta}(\BS, T) + 
                   \cfrac{1}{\rho_0}~\BS:\tilde{\BE}(\BS, T)~.
    \]
    Therefore,
    \[
       \Partial{\hat{e}(\BE,T)}{T} = \Partial{\hat{\psi}}{T}
          + \hat{\eta}(\BE, T) + T~\Partial{\hat{\eta}}{T}
    \]
    and
    \[
       \Partial{\tilde{e}(\BS,T)}{T} = - \Partial{\tilde{g}}{T} 
         + \tilde{\eta}(\BS,T) + T~\Partial{\tilde{\eta}}{T}
         + \cfrac{1}{\rho_0}~\BS:\Partial{\tilde{\BE}}{T} ~.
    \]
    Also, recall that
    \[
      \hat{\eta}(\BE,T)  = -\Partial{\hat{\psi}}{T}  
      \qquad \implies \qquad
      \Partial{\hat{\eta}}{T}  = -\PPartial{\hat{\psi}}{T}  ~,
    \]
    \[
      \tilde{\eta}(\BS, T) =  \Partial{\tilde{g}}{T}  
      \qquad \implies \qquad
      \Partial{\tilde{\eta}}{T} =  \PPartial{\tilde{g}}{T}  ~,
    \]
    and
    \[
      \tilde{\BE}(\BS, T) = \rho_0~\Partial{\tilde{g}}{\BS}
      \qquad \implies \qquad
      \Partial{\tilde{\BE}}{T} = \rho_0~\PPartialA{\tilde{g}}{\BS}{T}~.
    \]
    Hence,
    \[
      \Partial{\hat{e}(\BE,T)}{T} =  T~\Partial{\hat{\eta}}{T} = 
       -T~\PPartial{\hat{\psi}}{T} 
    \]
    and
    \[
       \Partial{\tilde{e}(\BS,T)}{T} =  
         T~\Partial{\tilde{\eta}}{T} + 
         \cfrac{1}{\rho_0}~\BS:\Partial{\tilde{\BE}}{T} = 
         T~\PPartial{\tilde{g}}{T}  
         + \BS:\PPartialA{\tilde{g}}{\BS}{T}~.
    \]

    \item
    {%\Blue 
      For thermoelastic materials, show that the balance of energy
      equation 
      \[
         \rho~T~\dot{\eta} = - \Div{\Bq} + \rho~s 
      \]
      can be expressed as either
      \[
      \rho~C_v~\Tdot  = \Div{(\Bkappa\cdot\Grad{T})} + \rho~s 
           +\cfrac{\rho}{\rho_0}~T~\Partial{\hat{\BS}}{T}:\BEdot 
      \]
      or
      \[
      \rho~\left(C_p - \cfrac{1}{\rho_0}~\BS:\Partial{\tilde{\BE}}{T}\right)
        ~\Tdot 
          = \Div{(\Bkappa\cdot\Grad{T})} + \rho~s 
           -\cfrac{\rho}{\rho_0}~T~\Partial{\tilde{\BE}}{T}:\BSdot  
      \]
      where 
      \[
      C_v = \Partial{\hat{e}(\BE,T)}{T} 
      \qquad \text{and} \qquad
      C_p = \Partial{\tilde{e}(\BS,T)}{T} ~.
      \]
    }
    If the independent variables are $\BE$ and $T$, then
    \[
       \eta = \hat{\eta}(\BE, T) \qquad \implies \qquad
       \Etadot = \Partial{\hat{\eta}}{\BE}:\BEdot + 
           \Partial{\hat{\eta}}{T}~\Tdot ~.
    \]
    On the other hand, if we consider $\BS$ and $T$ to be the independent 
    variables
    \[
       \eta = \tilde{\eta}(\BS, T) \qquad \implies \qquad
       \Etadot = \Partial{\tilde{\eta}}{\BS}:\BSdot + 
           \Partial{\tilde{\eta}}{T}~\Tdot ~.
    \]
    Since
    \[
       \Partial{\hat{\eta}}{\BE} = -\cfrac{1}{\rho_0}~\Partial{\hat{\BS}}{T} 
       ~;~~
       \Partial{\hat{\eta}}{T} = \cfrac{C_v}{T} ~;~~
       \Partial{\tilde{\eta}}{\BS} = 
       \cfrac{1}{\rho_0}~\Partial{\tilde{\BE}}{T} ~;~~\text{and}~~
       \Partial{\tilde{\eta}}{T} = \cfrac{1}{T}\left(C_p - \cfrac{1}{\rho_0}
         \BS:\Partial{\tilde{\BE}}{T}\right)
    \]
    we have, either
    \[
       \Etadot = -\cfrac{1}{\rho_0}~\Partial{\hat{\BS}}{T}:\BEdot + 
           \cfrac{C_v}{T}~\Tdot 
    \]
    or
    \[
       \Etadot = \cfrac{1}{\rho_0}~\Partial{\tilde{\BE}}{T}:\BSdot + 
         \cfrac{1}{T}\left(C_p - \cfrac{1}{\rho_0}
         \BS:\Partial{\tilde{\BE}}{T}\right)~\Tdot ~.
    \]
    The equation for balance of energy in terms of the specific entropy is
    \[
       \rho~T~\dot{\eta} = - \Div{\Bq} + \rho~s ~.
    \]
    Using the two forms of $\Etadot$, we get two forms of the energy equation:
    \[
       -\cfrac{\rho}{\rho_0}~T~\Partial{\hat{\BS}}{T}:\BEdot + 
           \rho~C_v~\Tdot  = - \Div{\Bq} + \rho~s 
    \]
    and
    \[
       \cfrac{\rho}{\rho_0}~T~\Partial{\tilde{\BE}}{T}:\BSdot + 
         \rho~C_p~\Tdot  
        - \cfrac{\rho}{\rho_0}~\BS:\Partial{\tilde{\BE}}{T}~\Tdot 
           = - \Div{\Bq} + \rho~s ~.
    \]
    From Fourier's law of heat conduction
    \[
       \Bq = - \Bkappa\cdot\Grad{T} ~.
    \]
    Therefore, 
    \[
       -\cfrac{\rho}{\rho_0}~T~\Partial{\hat{\BS}}{T}:\BEdot + 
           \rho~C_v~\Tdot  = \Div{(\Bkappa\cdot\Grad{T})} + \rho~s 
    \]
    and
    \[
       \cfrac{\rho}{\rho_0}~T~\Partial{\tilde{\BE}}{T}:\BSdot + 
         \rho~C_p~\Tdot  
        - \cfrac{\rho}{\rho_0}~\BS:\Partial{\tilde{\BE}}{T}~\Tdot 
           = \Div{(\Bkappa\cdot\Grad{T})} + \rho~s ~.
    \]
    Rearranging, 
    \[ 
      \rho~C_v~\Tdot  = \Div{(\Bkappa\cdot\Grad{T})} + \rho~s 
           +\cfrac{\rho}{\rho_0}~T~\Partial{\hat{\BS}}{T}:\BEdot 
    \]
    or,
    \[
      \rho~\left(C_p - \cfrac{1}{\rho_0}~\BS:\Partial{\tilde{\BE}}{T}\right)
        ~\Tdot 
          = \Div{(\Bkappa\cdot\Grad{T})} + \rho~s 
           -\cfrac{\rho}{\rho_0}~T~\Partial{\tilde{\BE}}{T}:\BSdot  ~.
    \]

    \item
    {%\Blue 
      For thermoelastic materials, show that the specific heats are
      related by the relation
      \[
         C_p - C_v = \cfrac{1}{\rho_0}\left(\BS-T~\Partial{\hat{\BS}}{T}\right):
           \Partial{\tilde{\BE}}{T} ~.
      \]
    }
    Recall that
    \[
       C_v := \Partial{\hat{e}(\BE,T)}{T} =  T~\Partial{\hat{\eta}}{T} 
    \]
    and
    \[
       C_p := \Partial{\tilde{e}(\BS,T)}{T} =  
           T~\Partial{\tilde{\eta}}{T} + 
           \cfrac{1}{\rho_0}~\BS:\Partial{\tilde{\BE}}{T} ~.
    \]
    Therefore,
    \[
       C_p - C_v = T~\Partial{\tilde{\eta}}{T} 
           + \cfrac{1}{\rho_0}~\BS:\Partial{\tilde{\BE}}{T}  
           - T~\Partial{\hat{\eta}}{T} ~.
    \]
    Also recall that
    \[
       \eta = \hat{\eta}(\BE, T) = \tilde{\eta}(\BS, T) ~.
    \]
    Therefore, keeping $\BS$ constant while differentiating, we have
    \[
       \Partial{\tilde{\eta}}{T} = \Partial{\hat{\eta}}{\BE}:\Partial{\BE}{T} + 
           \Partial{\hat{\eta}}{T} ~.
    \]
    Noting that $\BE = \tilde{\BE}(\BS,T)$, and
    plugging back into the equation for the difference between the two
    specific heats, we have
    \[
       C_p - C_v = T~\Partial{\hat{\eta}}{\BE}:\Partial{\tilde{\BE}}{T} 
           + \cfrac{1}{\rho_0}~\BS:\Partial{\tilde{\BE}}{T}  ~.
    \]
    Recalling that
    \[
       \Partial{\hat{\eta}}{\BE} = - \cfrac{1}{\rho_0}~\Partial{\hat{\BS}}{T}
    \]
    we get
    \[
        C_p - C_v = \cfrac{1}{\rho_0}\left(\BS -T~\Partial{\hat{\BS}}{T}\right):
           \Partial{\tilde{\BE}}{T} ~.
    \]

    \item
    {%\Blue 
      For thermoelastic materials, show that the specific heats 
      can also be related by the equations
      \[
         C_p - C_v = \cfrac{1}{\rho_0}~\BS:\Partial{\BE}{T} + 
           \Partial{\BE}{T}:\left(\PPartialA{\psi}{\BE}{\BE}:
             \Partial{\BE}{T}\right) 
         = \cfrac{1}{\rho_0}~\BS:\Partial{\BE}{T} + 
           \cfrac{T}{\rho_0}~
            \Partial{\BE}{T}:\left(\Partial{\BS}{\BE}:\Partial{\BE}{T}\right)
          ~.
      \]
    }
    Recall that
    \[
       \BS = \rho_0~\Partial{\psi}{\BE} = \rho_0~\BfT(\BE(\BS,T),T)~.
    \]
    Recall the chain rule which states that if
    \[
       g(u,t) = f(x(u,t), y(u,t))
    \]
    then, if we keep $u$ fixed, the partial derivative of $g$ with respect
    to $t$ is given by
    \[
      \Partial{g}{t} = \Partial{f}{x}~\Partial{x}{t} + 
                       \Partial{f}{y}~\Partial{y}{t} ~.
    \]
    In our case, 
    \[
       u = \BS, ~~t = T, ~~g(\BS, T) = \BS, ~~x(\BS,T) = 
       \BE(\BS,T), ~~y(\BS,T) = T,~~ \text{and}~~ f = \rho_0~\BfT.  
    \]
    Hence, we have
    \[
       \BS = g(\BS, T) = f(\BE(\BS,T), T) = \rho_0~\BfT(\BE(\BS,T),T)~.
    \]
    Taking the derivative with respect to $T$ keeping $\BS$ constant, we have
    \[
       \Partial{g}{T} = \cancelto{0}{\Partial{\BS}{T}} = 
         \rho_0~\left[\Partial{\BfT}{\BE}:
         \Partial{\BE}{T} +\Partial{\BfT}{T}~\cancelto{1}{\Partial{T}{T}}\right]
    \]
    or,
    \[
       \Bzero = \Partial{\BfT}{\BE}:\Partial{\BE}{T} + \Partial{\BfT}{T}~.
    \]
    Now,
    \[
      \BfT = \Partial{\psi}{\BE}
      \qquad \implies \qquad
      \Partial{\BfT}{\BE} =  \PPartialA{\psi}{\BE}{\BE}
      \quad \text{and} \quad
      \Partial{\BfT}{T} =  \PPartialA{\psi}{T}{\BE} ~.
    \] 
    Therefore,
    \[
       \Bzero = \PPartialA{\psi}{\BE}{\BE}:\Partial{\BE}{T} + 
                \PPartialA{\psi}{T}{\BE}
              = \Partial{}{\BE}\left(\Partial{\psi}{\BE}\right):
                \Partial{\BE}{T} + 
                \Partial{}{T}\left(\Partial{\psi}{\BE}\right) ~.
    \]
    Again recall that, 
    \[
       \Partial{\psi}{\BE} = \cfrac{1}{\rho_0}~\BS ~.
    \]
    Plugging into the above, we get 
    \[
       \Bzero = \PPartialA{\psi}{\BE}{\BE}:\Partial{\BE}{T} + 
                \cfrac{1}{\rho_0}~\Partial{\BS}{T} =
           \cfrac{1}{\rho_0}~\Partial{\BS}{\BE}:\Partial{\BE}{T} + 
                \cfrac{1}{\rho_0}~\Partial{\BS}{T} ~.
    \]
    Therefore, we get the following relation for $\partial \BS/\partial T$:
    \[
       \Partial{\BS}{T} 
          = - \rho_0~\PPartialA{\psi}{\BE}{\BE}:\Partial{\BE}{T} 
          = - \Partial{\BS}{\BE}:\Partial{\BE}{T} ~.
    \]
    Recall that
    \[
       C_p - C_v = \cfrac{1}{\rho_0}\left(\BS-T~\Partial{\BS}{T}\right):
           \Partial{\BE}{T} ~.
    \]
    Plugging in the expressions for  $\partial \BS/\partial T$ we get:
    \[
       C_p - C_v = \cfrac{1}{\rho_0}
          \left(\BS+T~\rho_0~\PPartialA{\psi}{\BE}{\BE}:\Partial{\BE}{T} 
           \right): \Partial{\BE}{T} 
         = \cfrac{1}{\rho_0}
           \left(\BS+T~\Partial{\BS}{\BE}:\Partial{\BE}{T}\right)
           :\Partial{\BE}{T} ~.
    \]
    Therefore,
    \[
       C_p - C_v = \cfrac{1}{\rho_0}~\BS:\Partial{\BE}{T} + 
          T~\left(\PPartialA{\psi}{\BE}{\BE}:\Partial{\BE}{T} 
           \right): \Partial{\BE}{T} 
         = \cfrac{1}{\rho_0}~\BS:\Partial{\BE}{T} + 
           \cfrac{T}{\rho_0}~\left(\Partial{\BS}{\BE}:\Partial{\BE}{T}\right)
           :\Partial{\BE}{T} ~.
    \]
    Using the identity $ (\SfA:\BB):\BC = \BC:(\SfA:\BB)$, we have
    \[
       C_p - C_v = \cfrac{1}{\rho_0}~\BS:\Partial{\BE}{T} + 
          T~\Partial{\BE}{T}:\left(\PPartialA{\psi}{\BE}{\BE}:\Partial{\BE}{T} 
           \right)
         = \cfrac{1}{\rho_0}~\BS:\Partial{\BE}{T} + 
           \cfrac{T}{\rho_0}~
            \Partial{\BE}{T}:\left(\Partial{\BS}{\BE}:\Partial{\BE}{T}\right)
            ~.
    \]
   
    \item
    {%\Blue 
      Consider an isotropic thermoelastic material that has a
      constant coefficient of thermal expansion and which follows the
      St-Venant Kirchhoff model, i.e,
      \[
         \Balpha_E = \alpha~\Bone \qquad\text{and}\qquad
         \SfC = \lambda~\Dyad{\Bone}{\Bone} + 2\mu~\SfI
      \]
      where $\alpha$ is the coefficient of thermal expansion and
      $3~\lambda = 3~K - 2~\mu$ where $K, \mu$ are the bulk and shear 
      moduli, respectively. 

      Show that the specific heats related by the equation
      \[
         C_p - C_v = \cfrac{1}{\rho_0}\left[\alpha~\Tr{\BS} + 
           9~\alpha^2~K~T\right]~.
      \]
    }
    Recall that,
    \[ 
    C_p - C_v = \cfrac{1}{\rho_0}~\BS:\Balpha_E + 
           \cfrac{T}{\rho_0}~\Balpha_E:\SfC:\Balpha_E ~.
    \]
    Plugging the expressions of $\Balpha_E$ and $\SfC$ into the above
    equation, we have
    \[ 
      \begin{aligned}
        C_p - C_v & = \cfrac{1}{\rho_0}~\BS:(\alpha~\Bone) + 
               \cfrac{T}{\rho_0}~(\alpha~\Bone):
               (\lambda~\Dyad{\Bone}{\Bone} + 2\mu~\SfI):
               (\alpha~\Bone) \\
         & = \cfrac{\alpha}{\rho_0}~\Tr{\BS} + 
               \cfrac{\alpha^2~T}{\rho_0}~\Bone:
               (\lambda~\Dyad{\Bone}{\Bone} + 2\mu~\SfI):
               \Bone \\
         & = \cfrac{\alpha}{\rho_0}~\Tr{\BS} + 
               \cfrac{\alpha^2~T}{\rho_0}~\Bone:
               (\lambda~\Tr{\Bone}~\Bone + 2\mu~\Bone)\\
         & = \cfrac{\alpha}{\rho_0}~\Tr{\BS} + 
               \cfrac{\alpha^2~T}{\rho_0}~
               (3~\lambda~\Tr{\Bone} + 2\mu~\Tr{\Bone})\\
         & = \cfrac{\alpha}{\rho_0}~\Tr{\BS} + 
               \cfrac{3~\alpha^2~T}{\rho_0}~
               (3~\lambda + 2\mu)\\
         & = \cfrac{\alpha~\Tr{\BS}}{\rho_0} + 
               \cfrac{9~\alpha^2~K~T}{\rho_0}~.
      \end{aligned}
    \]
    Therefore, 
    \[
       C_p - C_v = \cfrac{1}{\rho_0}\left[\alpha~\Tr{\BS} + 
           9~\alpha^2~K~T\right]~.
    \]
  \end{enumerate}
  }

\end{document}